\documentclass[pra,aps,twocolumn,showpacs,nofootinbib]{revtex4}
\usepackage{amsmath,graphicx,epsfig}

\def\prn#1{{\left(#1\right)}}

\def\ts#1{{_{\mbox{\scriptsize #1}}}}

\def\dbyd#1#2{{\frac{d #1}{d #2}}}

\def\fig_width{3. in} 
\draft

\newlength{\defbaselineskip}
\newcommand{\setlinespacing}[1]%
           {\setlength{\baselineskip}{#1 \defbaselineskip}}

\begin{document}

\title{Magnetometric sensitivity optimization for nonlinear optical rotation with frequency-modulated light: rubidium D2 line} 

\author{D. F. Jackson Kimball}
\email{derek.jacksonkimball@csueastbay.edu}
\author{L. R. Jacome}
\author{Srikanth Guttikonda}
\author{Eric J. Bahr}
\author{Lok Fai Chan}
\affiliation{Department of Physics, California State University --
East Bay, Hayward, California 94542-3084}

\date{\today}



\begin{abstract}

Atomic spin polarization of alkali atoms in the ground state can survive thousands of collisions with paraffin-coated cell walls.  The resulting long spin-relaxation times achieved in evacuated, paraffin-coated cells enable precise measurement of atomic spin precession and energy shifts of ground-state Zeeman sublevels.  In the present work, nonlinear magneto-optical rotation with frequency-modulated light (FM NMOR) is used to measure magnetic-field-induced spin precession for rubidium atoms contained in a paraffin-coated cell.  The magnetometric sensitivity of FM NMOR for the rubidium D2 line is studied as a function of light power, detuning, frequency-modulation amplitude, and rubidium vapor density.  For a 5-cm diameter cell at temperature $T \approx 35^\circ$C, the optimal shot-noise-projected magnetometric sensitivity is found to be $2 \times 10^{-11}~{\rm G/\sqrt{ Hz }}$ (corresponding to a sensitivity to spin precession frequency of $\approx 10~{\rm \mu Hz/\sqrt{Hz}}$ or a sensitivity to Zeeman sublevel shifts of $\approx 4 \times 10^{-20}~{\rm eV/\sqrt{Hz}}$).

\end{abstract}


%
%



\maketitle

\section{Introduction}

One of the earliest applications of techniques to spin-polarize alkali atoms using optical pumping \cite{Kas50,Bro52,Haw53} was to the problem of measuring magnetic fields \cite{Deh57a,Bel57,Bel61}.  The component of atomic spin polarization transverse to the direction of a magnetic field of magnitude $B$ precesses at the Larmor frequency $\Omega_L = g_F \mu_0 B$, where $g_F$ is the Land\'e g-factor for the atomic state and $\mu_0$ is the Bohr magneton.  A measurement of $\Omega_L$ therefore directly determines the value of $B$.  The spin-projection-noise-limited (shot-noise-limited) sensitivity $\delta B\ts{SNL}$ of the polarized atomic sample to magnetic fields is determined by the total number of atoms $N$ and the relaxation rate $\gamma\ts{rel}$ of the atomic polarization for measurement times $\tau \gg \gamma\ts{rel}^{-1}$ \cite{Auz06}:
\begin{align}
\delta B\ts{SNL} \approx \frac{1}{g_F \mu_0} \sqrt{ \frac{\gamma\ts{rel}}{N \tau} }~.
\label{Eq:FundamentalSensitivity-atoms}
\end{align}
As can be seen from Eq.~\eqref{Eq:FundamentalSensitivity-atoms}, one route to improving magnetometric sensitivity is to reduce $\gamma\ts{rel}$.  In the earliest optical pumping experiments (see Ref.~\cite{Hap72} for an extensive review), $\gamma\ts{rel}$ was dominated by relaxation due to wall collisions.  Two methods were developed to address the problem of wall relaxation in alkali vapor cells: filling the cell with a buffer gas \cite{Bro55,Coh57,Deh57b,Ska57,Har58} and coating the cell walls with paraffin \cite{Rob58,Bou66}.  In the present work, we follow the latter approach and employ an evacuated paraffin-coated cell \cite{Ale02}, a sphere 5~cm in diameter, for which $\gamma\ts{rel}$ due to spin-exchange and wall collisions is $\approx 2\pi \times 1~{\rm Hz}$ at room temperature ($T \approx 20^\circ$C).

\begin{figure}[b]
\centerline{\psfig{figure=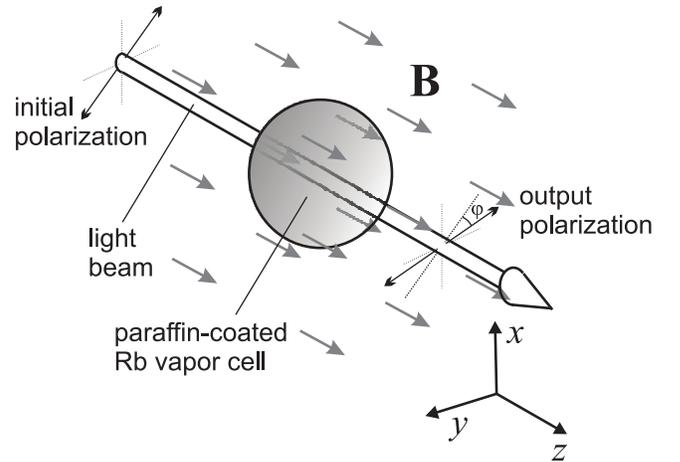,width=3.35 in}} \caption{Experimental geometry (known as the Faraday geometry) for measurement of magneto-optical rotation, where in our case the atomic medium is a sample of Rb atoms contained in a paraffin-coated cell.  The light propagates along the magnetic field $\bf{B}$ (which defines the longitudinal direction, $\hat{z}$).  The light is initially linearly polarized along an axis at $45^\circ$ to the $x$ and $y$ axes, and the plane of light polarization is rotated by an angle $\varphi$ at the output of the medium.} \label{Fig:Geometry}
\end{figure}

In order to realize the potential shot-noise-limited sensitivity of an optical pumping magnetometer described by Eq.~\eqref{Eq:FundamentalSensitivity-atoms}, an efficient method of detecting atomic spin precession is required.  In parallel with the extensive research on optical pumping, numerous experimental and theoretical studies of nonlinear magneto-optical rotation (NMOR, also known as nonlinear Faraday rotation, reviewed in Refs.~\cite{Bud02review,Ale05}) were being carried out.  The effect occurs when linearly polarized light propagates through an atomic medium along the direction of an applied magnetic field (Fig.~\ref{Fig:Geometry}).  When the light is near-resonant with an atomic transition, and of sufficient power to perturb the equilibrium population of atomic states, light-power-dependent rotation of the plane of light polarization is observed.  Research on optical pumping and studies of NMOR often overlapped, but for the present investigation the most important intersection of these two lines of inquiry was the discovery of narrow ($\sim 1~{\rm Hz}$) NMOR resonances in paraffin-coated cells \cite{Kan95,Bud98}.  These narrow NMOR resonances are related to optical pumping of long-lived ground-state atomic spin polarization.  A detailed study \cite{Bud00sens} of the magnetometric sensitivity of NMOR in a paraffin-coated rubidium (Rb) cell demonstrated that with proper choice of laser light power and detuning, it was possible, in principle, to achieve sensitivities close to the fundamental limit described by Eq.~\eqref{Eq:FundamentalSensitivity-atoms}, establishing NMOR as an highly efficient method of probing spin precession.

Shortly after the discovery of narrow NMOR resonances in paraffin-coated cells, it was realized that there were considerable practical advantages for atomic magnetometry if modulated light was used in the experimental scheme.  The first implementation of this idea \cite{Bud02} employed a single, frequency-modulated light beam for optical pumping and detection of NMOR resonances (FM NMOR).  (Although this initial work was inspired by frequency-modulation techniques employed in measurements of parity-violating optical rotation and linear Faraday rotation \cite{Bar78,Bar88}, the technique bears a resemblance to the early work of Bell and Bloom \cite{Bel61}, in which the intensity of a circularly polarized light beam was modulated synchronously with the Larmor precession of alkali atoms.)  The advantages of using frequency-modulated light are twofold.  First, narrow ($\sim 1~{\rm Hz}$) FM NMOR resonances appear at magnetic fields where the modulation frequency $\Omega_m$ coincides with a multiple of $\Omega_L$ \cite{Yas03}, considerably extending the dynamic range of an NMOR-based atomic magnetometer \cite{Aco06}.  Second, noise and systematic effects associated with spurious rotations are greatly reduced in the FM NMOR scheme because most sources of spurious rotation do not share the sharp spectral dependence of the atomic resonances (and thus do not produce significant optical rotation at the modulation frequency).  Furthermore, frequency modulation moves the detected signal away from $1/f$ noise.

The investigation of FM NMOR in paraffin-coated cells is but one branch of the increasingly vibrant and diverse field of atomic magnetometry \cite{Ale96,Bud07}.  Nonlinear magneto-optical rotation with amplitude-modulated light (AM NMOR) in paraffin-coated cells has been investigated, in both single laser beam \cite{Gaw06,Bal06,Pus07,Pus08} and two-beam pump/probe arrangements \cite{Hig06}.  Atomic magnetometry using alkali vapor cells filled with buffer gas has been extensively studied \cite{Sau00,Sta01,Aff02,Nov02,Nov05,Wei06,Dom07}, and applied, for example, to biomedical measurements \cite{Wei04}.

One of the most significant developments in atomic magnetometry in recent years has been the invention of the spin-exchange-relaxation-free (SERF) magnetometer \cite{All02,Kom03,Sha07,Led08}.  A SERF magnetometer operates under conditions of high alkali vapor density where the spin-exchange rate $\gamma\ts{se}$ is much faster than the Larmor frequency $\Omega_L$.  In this regime, rapid spin-exchange causes atomic polarization in the two ground-state hyperfine levels to become strongly correlated \cite{Hap77}, and spin-exchange is effectively eliminated as a source of relaxation.  SERF magnetometers have shot-noise-projected sensitivities of $\sim 10^{-14}~{\rm G/\sqrt{Hz}}$ and have achieved sensitivities of $\sim 10^{-11}~{\rm G/\sqrt{Hz}}$ in practice \cite{Kom03}, although their dynamic range is limited by the condition $\Omega_L \ll \gamma\ts{se}$.  Compared to SERF magnetometers, alkali vapor magnetometers based on FM NMOR in paraffin-coated cells have the practical advantage of significantly greater dynamic range.

Atomic magnetometers based on FM NMOR have already been applied to measurements of nuclear magnetism \cite{Yas04,Cra08,Led09}, magnetic resonance imaging (MRI) \cite{Xu06a,Xu06b,Xu08}, geophysical field measurements \cite{Aco06}, and magnetic particle detection \cite{Xu06c}.  A self-oscillating FM NMOR atomic magnetometer has been constructed \cite{Sch05} and FM NMOR in paraffin-coated cells of diameter $\approx 3~{\rm mm}$ has been observed \cite{Bal06} in an ongoing effort to develop chip-scale (dimensions $\sim 1~{\rm mm}$) magnetometers \cite{Kna06}.  The technique of FM NMOR has also enabled selective creation and detection of high-order atomic polarization moments \cite{Yas03,Pus06,Aco08}. In spite of the growing body of work involving FM NMOR in paraffin-coated cells, the essential question of under what conditions is the optimal magnetometric sensitivity achieved has yet to be addressed.  This is the subject of the present study, in which a systematic optimization of the magnetometric sensitivity of FM NMOR with respect to light power, light detuning, and modulation amplitude is carried out for the Rb D2 line.  The dependence of the sensitivity on cell temperature and Rb vapor density is also investigated.  Several details of the FM NMOR spectrum are explained and applications of FM NMOR techniques to problems in fundamental physics are briefly discussed in the conclusion.

\section{Experimental setup}

In the present set of experiments, as in past studies of FM NMOR using a single-beam arrangement \cite{Bud02,Yas03,Mal04,Aco06,Pus06b,Pus06c}, we employ the Faraday geometry (Fig.~\ref{Fig:Geometry}) where a linearly polarized laser beam propagates along the direction ($\hat{z}$) of a magnetic field $\mathbf{B}=B_z\hat{z}$.  The resonant (or near-resonant) interaction of the light beam with a sample of Rb atoms generates ground-state atomic polarization via optical pumping. The optically pumped ground-state atomic polarization evolves in the presence of the magnetic field $\mathbf{B}$: at sufficiently low light power, the evolution is simply Larmor precession of the aligned atomic spins, while at higher light power, ac Stark shifts of the ground-state Zeeman sublevels due to the optical electric field in combination with the Zeeman shifts due to $\mathbf{B}$ orient the atomic spins along $\hat{z}$ (alignment-to-orientation conversion, AOC \cite{Bud00aoc}).  Interaction of evolved ground-state atomic polarization with the laser light causes the plane of light polarization at the output of the vapor to rotate.

\begin{figure*}
\centerline{\psfig{figure=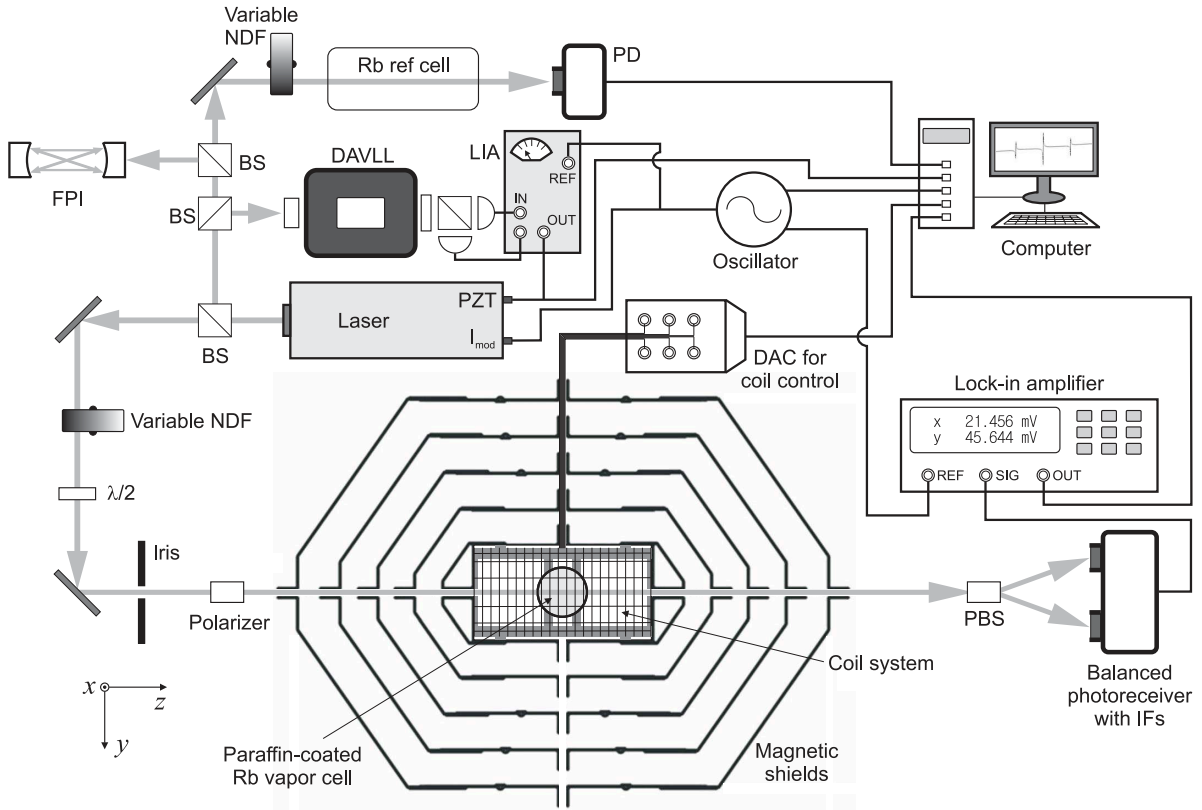,width=6.2 in}} \caption{Schematic diagram of the experimental setup.  FPI~=~Fabry-Perot interferometer, NDF~=~neutral density filter, PD~=~photodiode, BS~=~50/50 beam splitter, DAVLL~=~Dichroic Atomic Vapor Laser Lock system (see Refs.~\cite{Cor98,Yas00}), LIA~=~lock-in amplifier (analog), PZT~=~analog control input for PZT affecting extended-cavity diode laser feedback grating angle, ${\rm I\ts{mod}}$~=~analog control input for modulation of diode laser current, DAC~=~digital-to-analog converter, PBS~=~polarizing beam splitter (Wollaston), IF~=~interference filter.} \label{Fig:ExptSetup}
\end{figure*}

A schematic diagram of the experimental setup is shown in Fig.~\ref{Fig:ExptSetup}.  A tunable extended-cavity diode laser (Toptica DL100) is used to generate light at 780 nm resonant with the D2 transition for Rb ($^2S_{1/2}~\rightarrow~^2P_{3/2}$).  A computer-controlled oscillator (the programmable oscillator of the Signal Recovery model 7265 digital lock-in amplifier) sinusoidally modulates the diode laser current (${\rm I\ts{mod}}$) at a frequency $\Omega_m$.  The principal effect of the current modulation is frequency modulation of the laser light at $\Omega_m$, amplitude modulation of the laser light is less than 1\% of total power for the maximum modulation amplitude employed in our experiments.  The central frequency $\omega_0$ of the laser light is voltage-controlled using the feedback grating's piezo element, and can be scanned by computer or stabilized to a specific frequency in the Rb spectrum using feedback from the demodulated (via an analog lock-in amplifier, EG\&G PARC Model 5101) output signal of a temperature-stabilized dichroic atomic vapor laser lock (DAVLL) system \cite{Cor98,Yas00} for Rb.  A Fabry-Perot interferometer is monitored to ensure the laser light is single mode and to calibrate the modulation amplitude $\Delta \omega$.  The transmission spectrum through an uncoated Rb reference cell (natural isotopic mixture) is measured with a photodiode fitted with a 780-nm central wavelength interference filter (10~nm bandwidth) and recorded by computer (the transmission signal is demodulated with a lock-in amplifier when the laser current is modulated).  The laser light power through the reference cell is reduced with a neutral density filter (NDF) to a level sufficiently low ($\sim 10~{\rm \mu W}$) so that nonlinear optical effects distorting the transmission spectrum can be ignored.

A spherical paraffin-coated vapor cell, containing a natural isotopic mixture of Rb, is mounted inside a frame manufactured of HDPE (High Density Polyethylene). The frame is fit inside the innermost layer of a five-layer magnetic shield (manufactured by Amuneal Inc.) made of a 1-mm thick high-permeability alloy, annealed in a hydrogen atmosphere.  Each layer of the shield consists of a cylindrical center piece and two removable end caps.  The layers of the shield are spaced by styrofoam (polymerized in place).  Four ports for access to the inside of the shields are available on the cylindrical pieces and one port is available on each end cap.  The shielding factor of the entire five-layer magnetic shield system was measured to be better than $10^7$ \cite{Xu06a}.  A system of six separate coils are wound in grooves cut into the frame mounted inside the innermost layer of the shield.  The system of coils was designed to provide, over the volume of the paraffin-coated Rb vapor cell, uniform magnetic fields in three orthogonal directions ($B_x$, $B_y$, and $B_z$), linear magnetic field gradients in two directions ($\dbyd{B_x}{x}$, $\dbyd{B_z}{z}$), and a quadratic gradient along the shield axis ($\dbyd{^2B_z}{z^2}$).  Based on computer modeling (using the Amperes program from Integrated Engineering Software Inc.), the estimated uniformity of the magnetic fields and linearity/quadracity of the field gradients generated by the coil system is at a part per thousand over the cell volume for typical applied currents.  It should be noted that effects of uncompensated magnetic field gradients on the FM NMOR resonance width and amplitude are significantly reduced by motional averaging \cite{Pus06b} (effects are quadratic in the the magnitude of the gradient). The coils are in series with a set of ultra-stable, low temperature coefficient (low TC) resistors (Caddock Type USF 200 Series, zero nominal TC with TC $\lesssim 2~{\rm ppm/K}$), and voltages for the coils are computer generated with a digital-to-analog-converter (DAC, National Instruments PCI-6733).  For work requiring lower noise and superior stability, the voltage for the $B_z$ coil can also be supplied by a precision DC voltage source (Krohn-Hite Model 523 calibrator, stability $\pm 1~{\rm ppm}$).

The laser light beam that passes through the paraffin-coated Rb vapor cell first travels through a variable NDF and $\lambda/2$ plate, enabling control of the laser light power. The beam is apertured with an iris (resulting in a laser beam diameter of $\approx 2~{\rm mm}$) before passing through an antireflection-coated Glan Thomson linear polarizer (calcite, extinction ratio $5 \times 10^5 : 1$).  After exiting the paraffin-coated vapor cell, the beam is analyzed by a polarimeter consisting of a Wollaston prism polarizing beamsplitter (calcite, extinction ratio $10^5 : 1$) whose output rays are detected with a balanced photoreceiver (New Focus Model 2307) fitted with interference filters centered at 780~nm (bandwidth $\pm 10~{\rm nm}$).  The signal from the balanced photoreceiver is sent to the input of a digital lock-in amplifier (Signal Recovery model 7265).  The reference signal is the lock-in amplifier's internal oscillator that drives the current modulation of the diode laser.  The in-phase and quadrature components of the demodulated signal are recorded by computer (using routines written in LabVIEW).

\section{Magnetic field resonances}
\label{Sec:MagFieldRes}

\begin{figure}
\centerline{\psfig{figure=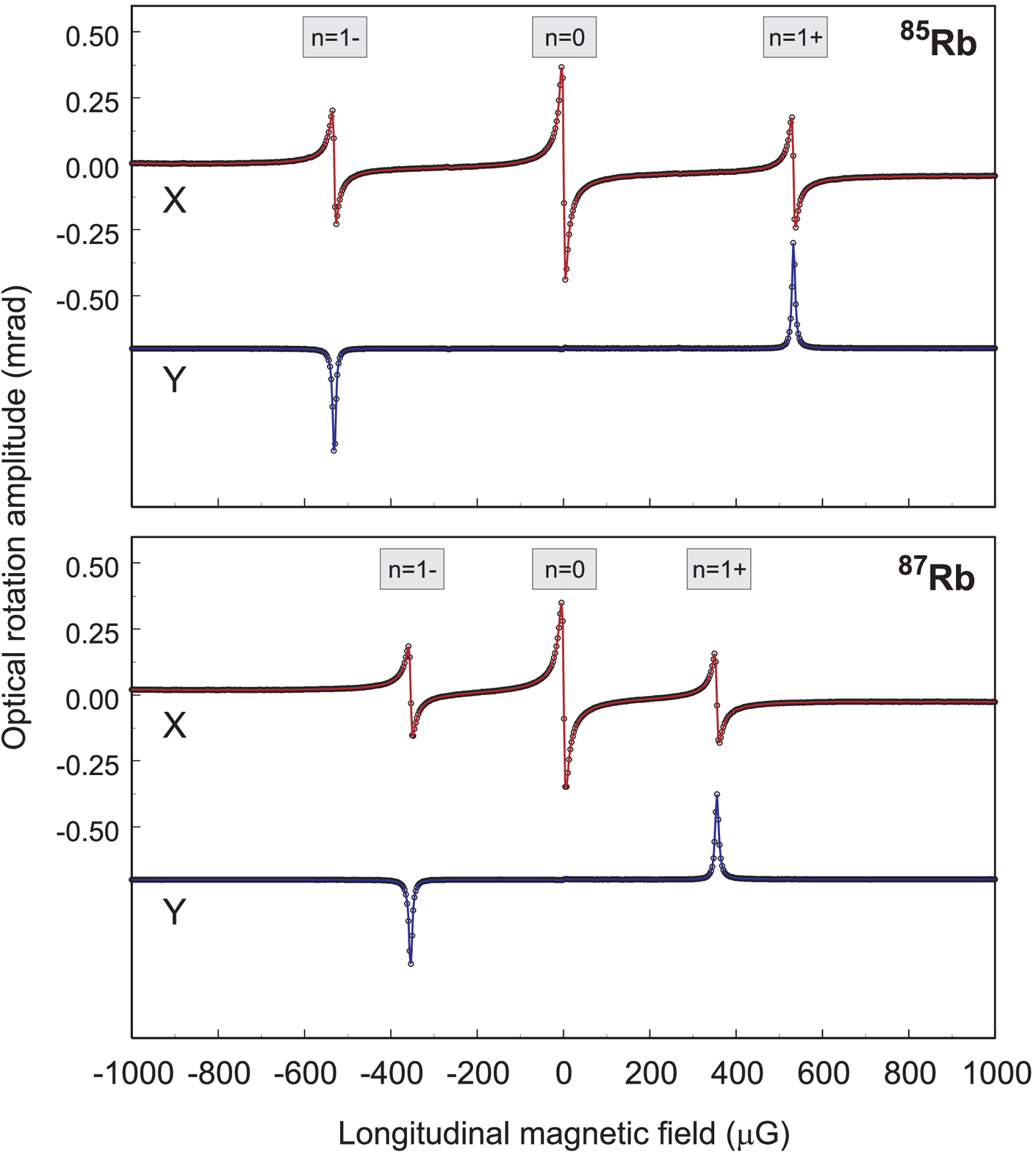,width=3.35 in}} \caption{Nonlinear magneto-optical rotation amplitude as a function of longitudinal magnetic field ($B_z$, along the direction of light propagation), demodulated at the first harmonic of $\Omega_m$.  The upper plot shows the $B_z$-dependence of the in-phase (X, data offset above) and out-of-phase (Y, data offset below) FM NMOR signal amplitudes when the laser is tuned to the high-frequency side of the Doppler-broadened $F=3 \rightarrow F'$ component of the $^{85}$Rb D2 transition.  The lower plot shows the $B_z$-dependence of the FM NMOR signal amplitudes when the laser is tuned to the high-frequency side of the Doppler-broadened $F=2 \rightarrow F'$ component of the $^{87}$Rb D2 transition.  The light power is 20~$\mu$W, the modulation amplitude $\Delta \omega = 65~{\rm MHz}$, the modulation frequency $\Omega_m = 500~{\rm Hz}$, and the cell temperature was $T = 20.5^\circ{\rm C}$ for which the Rb vapor density was measured to be $4 \times 10^9~{\rm atoms/cm^3}$ by fitting a low light power ($\approx 1~{\rm \mu W}$) absorption spectrum to a Voigt profile.  The FM NMOR resonances at $B_z=0$ are denoted the $n=0$ resonances, and the resonances at $B_z = \pm\Omega_m/(2g_F\mu_0)$ are denoted the $n=1\pm $ resonances (where $\Omega_m = 2\Omega_L$).} \label{Fig:BzScan}
\end{figure}

Figure~\ref{Fig:BzScan} shows the dependence of the optical rotation amplitude (demodulated by the lock-in amplifier) as a function of longitudinal magnetic field $B_z$ for two different laser light detunings: the upper plot shows the result of a magnetic field scan when the laser is tuned $\approx 250~{\rm MHz}$ to the high-frequency side of the Doppler-broadened $F=3 \rightarrow F'$ component of the $^{85}$Rb D2 transition, and the lower plot shows a magnetic field scan with the laser tuned $\approx 250~{\rm MHz}$ to the high-frequency side of the Doppler-broadened $F=2 \rightarrow F'$ component of the $^{87}$Rb D2 transition.  Prominent resonances in the magnetic field dependence of the optical rotation amplitude measured at the first harmonic of $\Omega_m$ are observed when
\begin{align}
n \Omega_m = 2 \Omega_L~,
\label{Eq:MagFieldResonanceCondition}
\end{align}
where $n=0,1$. (Smaller amplitude resonances in the FM NMOR signal demodulated at the first harmonic of $\Omega_m$ with $n > 1$ can be observed due to non-sinusoidal modulation of the light-atom interaction probability \cite{Bud02,Mal04} and misalignment between $\mathbf{B}$ and the wave vector $\mathbf{k}$ of the light beam \cite{Pus06c}; larger amplitude resonances with $n > 1$  can be observed when the FM NMOR signal is demodulated at higher harmonics of $\Omega_m$ \cite{Bud02,Mal04}.)  Because of the different Land\'{e} g-factors for $^{87}$Rb and $^{85}$Rb, the FM NMOR magnetic field resonances occur at different magnetic fields for the two isotopes. The magnetic field resonances described by Eq.~\eqref{Eq:MagFieldResonanceCondition} and observed in Fig.~\ref{Fig:BzScan} are related to Larmor precession of ground-state atomic alignment (the quadrupole, or $\kappa=2$, multipole moment of the angular momentum distribution, see Ref.~\cite{Bud02review} and references therein). (Additional resonances related to high-order ($\kappa > 2$) atomic polarization moments satisfying the resonance condition $n \Omega_m = \kappa \Omega_L$ can be observed at sufficiently high light power for transitions from hyperfine levels which can support such polarization moments ($F \geq \kappa/2$), see Refs.~\cite{Yas03,Pus06,Aco08}.)  The distribution of the angular momenta of a sample of aligned atoms has a preferred axis but no preferred direction, and consequently an aligned atomic vapor has different indices of refraction for light polarized parallel and orthogonal to the alignment axis.  Optical pumping initially creates alignment along the axis of linear polarization of the light, but due to Larmor precession of the atomic spins about the magnetic field, the alignment axis rotates about $\hat{z}$ at $\Omega_L$ and so the optical properties of the atomic sample are modulated at $2\Omega_L$ (the factor of 2 comes from the symmetry of the atomic alignment).  At sufficiently high light powers there is both Larmor precession of atomic alignment and evolution of atomic polarization related to ac Stark shifts leading to alignment-to-orientation conversion (AOC, see Ref.~\cite{Bud00aoc}), which generates atomic orientation (the dipole, or $\kappa=1$, multipole moment of the angular momentum distribution) in the $\hat{z}$-direction (along $\mathbf{B}$).  In this section, for simplicity, we discuss the physical mechanisms causing the $n=0$ and $n=1$ resonances under low light power conditions where FM NMOR is due to alignment precession.  FM NMOR due to AOC, which is the critical effect under the experimental conditions where optimum magnetometric sensitivity is achieved, is discussed in Sec.~\ref{Sec:LightPower}.

The $n=0$ resonance occurs under the conditions where $\Omega_L \ll \Omega_m$ and optical rotation achieves a maximum amplitude when $\Omega_L \approx \gamma\ts{rel}$.  At $B_z=0$, there is no Larmor precession and therefore no rotation.  As $B_z$ departs from zero the atomic alignment axis rotates, but if $\Omega_L \lesssim \gamma\ts{rel}$, optically pumped alignment relaxes before a full period of rotation can be completed.  Therefore in this regime, the average atomic alignment in the cell has its axis tilted away from the axis of the incident light polarization by some angle $\phi \lesssim \pi/4$.  Because of the different indices of refraction parallel and orthogonal to the alignment axis, the light polarization axis at the output of the cell is rotated with respect to the incident light polarization axis.  The frequency modulation of the light near the Doppler-broadened atomic resonance causes the optical rotation at the output to acquire a periodic time dependence as the probability of the light-atom interaction is modulated.  For the $n=0$ resonance, it should be noted that frequency modulation for the ``pump interaction'' which creates the initial atomic alignment is not essential for the effect, whereas frequency modulation for the ``probe interaction'' which causes optical rotation is essential to generate the time-dependent signal at the first harmonic of $\Omega_m$.  As $B_z$ is increased so that $\Omega_L \gtrsim \gamma\ts{rel}$, the optical rotation amplitude decreases since the alignment is rotated by $\phi \gtrsim \pi/4$ after it is initially produced via optical pumping, and eventually the magneto-optical rotation averages to zero as the alignment precesses by more than a full period before relaxing (causing dephasing of the alignment for atoms optically pumped at different times).  The shape of the resonance is well-described by a dispersive Lorentzian profile, as discussed in detail in Ref.~\cite{Mal04}.  The time-dependent optical rotation for the $n=0$ resonance is in-phase with the modulation of the probe interaction, and defines the phase of what is denoted the X signal in our experiments.

\begin{figure}
\centerline{\psfig{figure=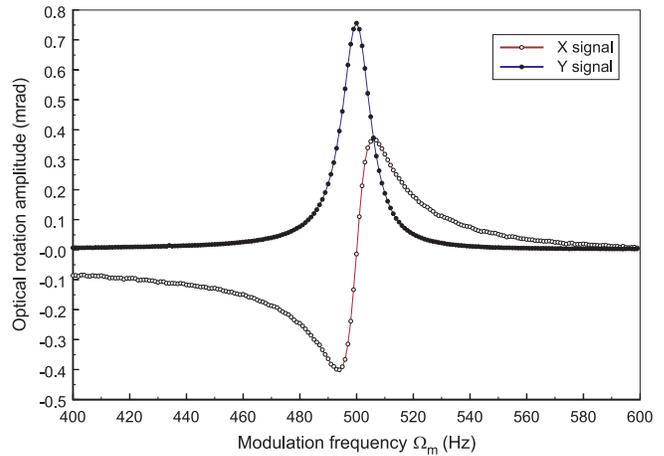,width=3.35 in}} \caption{FM NMOR ($n=1+$ resonance) amplitude as a function of modulation frequency $\Omega_m$ for longitudinal field $B_z=535.6~{\rm \mu G}$, laser tuned to the high-frequency side of the Doppler-broadened $F=3 \rightarrow F'$ component of the $^{85}$Rb D2 transition. The light power is 50~$\mu$W, the modulation amplitude $\Delta \omega = 65~{\rm MHz}$ and the cell temperature was $T = 25.6^\circ{\rm C}$ for which the Rb vapor density was measured to be $\approx 10^{10}~{\rm atoms/cm^3}$. X signal is offset from zero due to background rotation from the $n=0$ NMOR transit effect resonance, Y signal has negligible offset because there is no Y component for the $n=0$ transit effect resonance.} \label{Fig:ModFreqScan}
\end{figure}

The $n=1$ resonance occurs under the conditions where $\Omega_L \gg \gamma\ts{rel}$.  In the case of the $n=1$ resonance, in contrast to the $n=0$ case, the FM NMOR resonance can also be observed in the $\Omega_m$-dependence of the demodulated optical rotation signal (Fig.~\ref{Fig:ModFreqScan}). The physical mechanism giving rise to the $n=1$ resonance can be generally understood in analogy with a driven, damped harmonic oscillator: the atomic alignment produced by optical pumping naturally precesses at $\Omega_L$ due to the presence of $\mathbf{B}$, and when optical pumping (the driving term for the oscillator) is modulated at $2\Omega_L$ (the factor of 2 arises because of the two-fold symmetry of atomic alignment \cite{Yas03}), there is a resonant enhancement of the atomic alignment leading to an enhancement of the optical rotation signal.  As is the case with driven, damped harmonic oscillators, the phase of the time-dependent optical rotation acquires a dependence on the detuning of the drive frequency from the natural oscillation frequency ($2\Omega_L - \Omega_m$), thus signals are observed both in-phase (X signal) and out-of-phase (Y signal) with the modulation of the light-atom interaction probability.  While the X signal nominally \cite{endnote1,Bud02holeburning} crosses zero at $2\Omega_L = \Omega_m$, the Y signal is maximum when $2\Omega_L = \Omega_m$.  This is because when the optical pumping rate is synchronized with the atomic alignment precession rate, the axis of the atomic alignment is parallel with the light polarization at the periodic maxima in the modulated light-atom interaction probability --- when the atomic alignment axis is parallel with the light polarization no optical rotation is produced.  The maximum optical rotation occurs when the atomic alignment axis is rotated by an angle $\phi = \pi/4$ with respect to the light polarization, which on resonance ($2\Omega_L = \Omega_m$) causes optical rotation out-of-phase with the modulation of the light-atom interaction probability.

The combination of $n=0$ and $n=1$ resonances enable accurate determination of the magnetic field: the $n=0$ resonance provides a signal which can be used to determine the compensation fields required to set $\mathbf{B} = 0$ inside the shields with the coil system, and the $n=1$ resonances can be used to directly measure the dependence of $\Omega_L$ on the current applied to the $z$-coil.  In this way, the longitudinal magnetic field $B_z$ can be precisely calibrated.  Another method of magnetic field calibration is to measure the $n=1$ resonances for both Rb isotopes and use the known values of the Land\'e g-factors for the two isotopes to extract both the slope and offset terms describing the relationship between the $z$-coil current and $B_z$.  Conversely, precise, simultaneous measurement of $\Omega_L$ for both Rb isotopes can be used to measure the ratio of Land\'{e} g-factors for $^{85}$Rb and $^{87}$Rb and search for non-magnetic sources of spin-precession.

\section{Light-power dependence of FM NMOR spectra}
\label{Sec:LightPower}

\begin{figure*}
\centerline{\psfig{figure=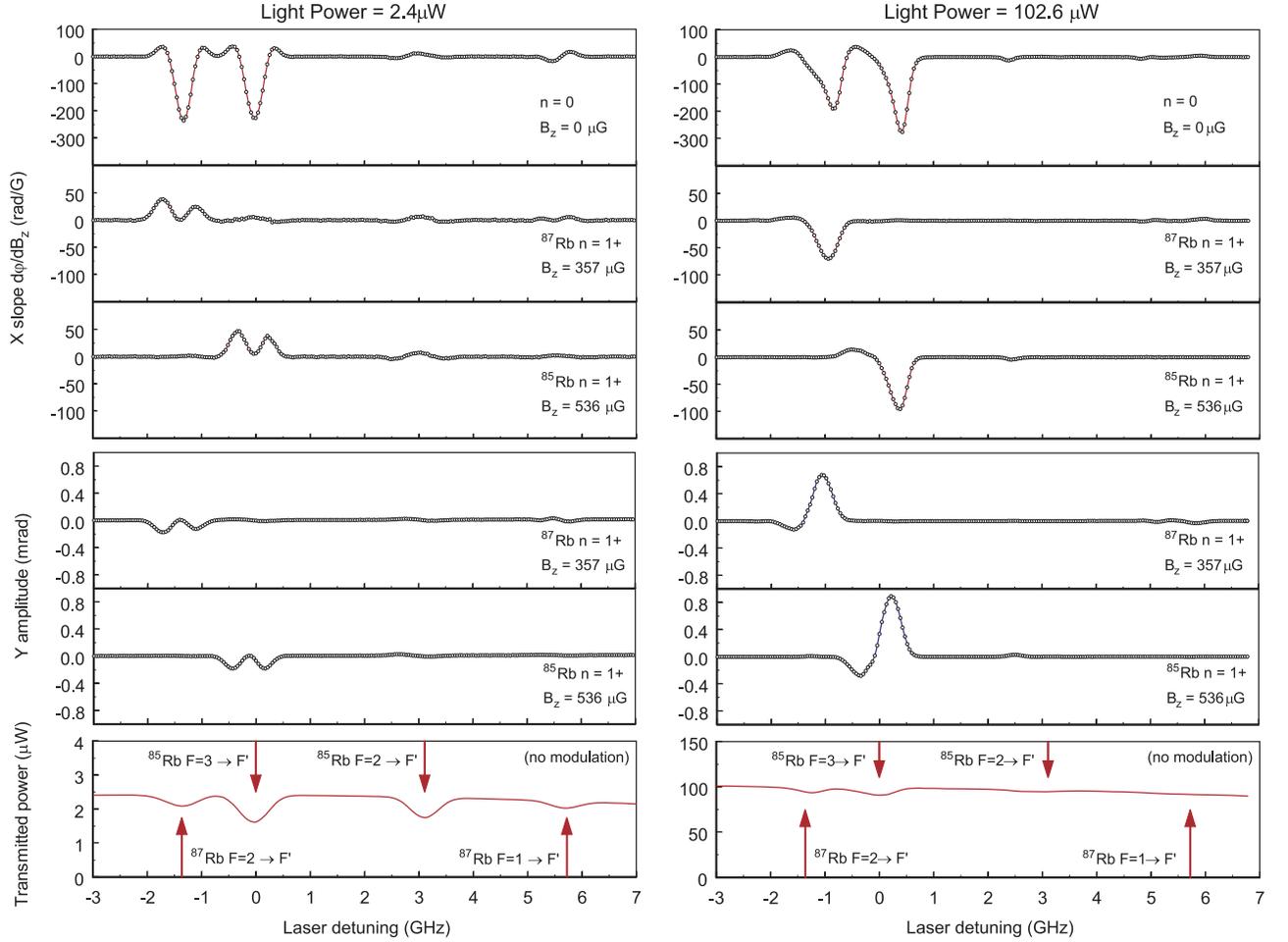,width=6.7 in}} \caption{Laser detuning dependence of FM NMOR signals for incident light power $2.4~{\rm \mu W}$ (left plots) and light power $102.6~{\rm \mu W}$ (right plots).  Laser modulation parameters are $\Omega_m = 500~{\rm Hz}$ and $\Delta \omega = 65~{\rm MHz}$, and the cell temperature $T = 25^\circ$C, corresponding to a vapor density of $\approx 10^{10}~{\rm atoms/cm^3}$.  The X signals are characterized by the derivative of the optical rotation amplitude with respect to longitudinal field ($d\varphi/dB_z$ in units of rad/G) and the Y signals are characterized by their amplitude (mrad).  Transmission spectra (with no frequency modulation) are shown at bottom, with arrows indicating the central frequencies of the various Doppler-broadened hyperfine components of the D2 transition.} \label{Fig:LowPowerHighPowerSpectra}
\end{figure*}

\begin{figure*}
\centerline{\psfig{figure=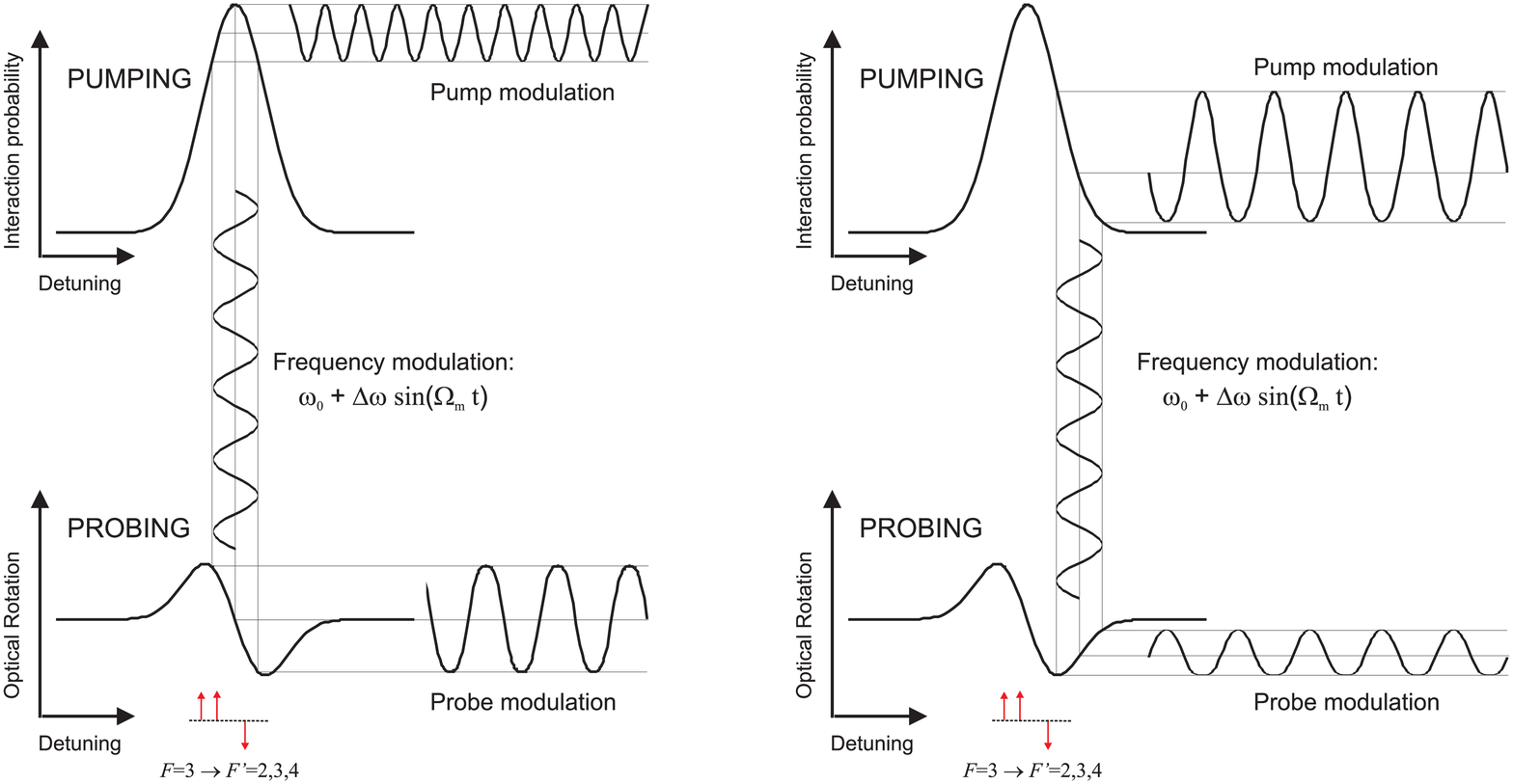,width=5.5 in}} \caption{Illustration of the time-dependent optical pumping and optical rotation generated via frequency modulation for low light power (where alignment-to-orientation can be neglected and optical rotation is of opposite sign for $F \rightarrow F+1$ transitions as compared to $F \rightarrow F-1, F$ transitions, see text).  The diagram on the left-hand side illustrates the pump and probe modulation when the center frequency of the laser light is tuned to the center of the Doppler-broadened $^{85}$Rb $F=3 \rightarrow F'$ transition, the diagram on the right-hand side illustrates pump and probe modulation when the laser light is detuned to the wing of the resonance.  Note that pump modulation is at the second harmonic of the frequency modulation when the laser light is tuned to the center of the resonance.  Arrows at the base of the probing interaction diagram indicate the approximate central frequencies and estimated relative contribution of different hyperfine components ($F \rightarrow F'$) to the overall Doppler-broadened optical rotation spectrum.} \label{Fig:Low-Power-PumpProbe-Modulation}
\end{figure*}

The spectral dependence of FM NMOR signals for the Rb D2 line for two representative light powers is shown in Fig.~\ref{Fig:LowPowerHighPowerSpectra}.  The X signal is characterized by the slope in rad/G, obtained by taking the difference between the X signals with
\begin{align}
B_z &= \pm \frac{1}{4} \frac{\gamma\ts{rel}(0)}{2g_F\mu_0} &(n=0) \nonumber \\
B_z &= \frac{\Omega_m}{2g_F\mu_0} \pm \frac{1}{4} \frac{\gamma\ts{rel}(0)}{2g_F\mu_0} &(n=1) \nonumber
\end{align}
and dividing by $\gamma\ts{rel}(0)/(4g_F\mu_0)$ on a point-by-point basis as the laser detuning is scanned through the Doppler-broadened Rb spectrum, where $\gamma\ts{rel}(0) \approx 1.3~{\rm Hz}$ is the width of the magnetic field resonance extrapolated to zero light power.  The Y signals are acquired on resonance $$B_z = \frac{\Omega_m}{2g_F\mu_0}~.$$  For reference, the transmission spectra without laser frequency modulation are displayed at the bottom of Fig.~\ref{Fig:LowPowerHighPowerSpectra} and the central frequencies of the Doppler broadened hyperfine components are indicated by arrows.

There is a significant difference between the shapes of the low-light-power ($P = 2.4~{\rm \mu W}$) $n=0$ and $n=1$ spectra: the maximum rotation for the $n=0$ X signal occurs close to the center of the Doppler-broadened $^{87}$Rb~$F=2 \rightarrow F'$  and $^{85}$Rb~$F=3 \rightarrow F'$ hyperfine components, whereas the maximum rotation for both the $n=1$ X and Y signals occurs on the wings of these components.  This is the result of differences between the roles of optical pumping and optical probing in the $n=0$ and $n=1$ cases.  In the $n=0$ case, the creation of atomic alignment via optical pumping does not require any special synchronization of the modulation frequency with $\Omega_L$ because the atomic alignment is nearly static; but in the $n=1$ case, atomic alignment is created only when the resonance condition $\Omega_m = 2 \Omega_L$ is satisfied so that macroscopic alignment precessing at $\Omega_L$ can be generated.  As Fig.~\ref{Fig:Low-Power-PumpProbe-Modulation} illustrates, when the laser is tuned to the center of the Doppler-broadened resonance, modulation of the light-atom interaction probability (pump modulation) occurs at the second harmonic of the modulation frequency.  While the frequency of pump modulation has no significant effect on optical pumping for the $n=0$ case, in the $n=1$ case the modulation is at $2\Omega_m = 4\Omega_L$ and is therefore non-optimal for optical pumping of macroscopic alignment precessing at $\Omega_L$.  This suppresses optical rotation at the center of the Doppler-broadened optical resonances for the $n=1$ case. On the other hand, when the laser is detuned to the wing of the Doppler-broadened optical resonance (Fig.~\ref{Fig:Low-Power-PumpProbe-Modulation}), frequency modulation modulates the pump interaction at the first harmonic of $\Omega_m$, which satisfies the optical pumping resonance condition for the $n=1$ case.  Note that because of the shape of optical rotation spectrum for the probe interaction (Fig.~\ref{Fig:Low-Power-PumpProbe-Modulation}, explained below), the probe interaction generates a strong signal at the first harmonic of $\Omega_m$ when tuned to the center of the Doppler-broadened optical resonance for both $n=0$ and $n=1$ resonances.

The shape of the low-light-power Doppler-broadened FM NMOR spectra for the $^{87}$Rb~$F=2 \rightarrow F'$ and $^{85}$Rb~$F=3 \rightarrow F'$ hyperfine components can be understood as follows.  At low light powers, the sign of magneto-optical rotation is opposite for $F \rightarrow F-1, F$ transitions as compared to $F \rightarrow F+1$ transitions \cite{Bud00sens,Bud00aoc,Bud02review}, as shown in Fig.~\ref{Fig:Low-Power-PumpProbe-Modulation}.  This is because the optically pumped alignment for $F \rightarrow F-1$ and $F \rightarrow F$ transitions corresponds to a ``dark state'' for which the light-atom interaction probability is reduced, while the optically pumped alignment for an $F \rightarrow F+1$ transition corresponds to a ``bright state'' for which the light-atom interaction probability is increased.  As a result, the Doppler-broadened optical rotation spectrum without modulation (for example, as studied in Refs.~\cite{Bud98,Bud00sens}) takes on a characteristic dispersive spectral dependence as illustrated in Fig.~\ref{Fig:Low-Power-PumpProbe-Modulation}.  With modulation, the maximum time-dependent rotation amplitude for the $n=0$ signal is obtained at the center of the Doppler-broadened profile where the slope of unmodulated rotation is steepest (in fact, the $n=0$ FM NMOR spectrum resembles the derivative of the unmodulated NMOR spectrum, as noted in Ref.~\cite{Bud02}).  The probe interaction generates optical rotation at the first harmonic of $\Omega_m$ both at the center and on the wings of the Doppler-broadened optical resonance.

Note that at both low and high light powers, and for both $n=0$ and $n=1$ resonances, the FM NMOR signals for the $^{87}$Rb $F = 1 \rightarrow F'$ and $^{85}$Rb $F = 2 \rightarrow F'$ hyperfine components are significantly smaller than the FM NMOR signals for the $^{87}$Rb~$F=2 \rightarrow F'$ and $^{85}$Rb~$F=3 \rightarrow F'$ hyperfine components.  This is also observed in NMOR experiments without frequency modulation \cite{Bud00sens,Bud02review}, and results from a combination of factors related to the optical pumping of atomic alignment and optical probing of the alignment precession.  The suppression of optical pumping and probing of atomic alignment for particular transitions is related to the excited-state hyperfine splitting: the NMOR amplitude for overlapping, Doppler-broadened hyperfine components of optical transitions is strongly dependent on the energy splitting between the $F'$ levels, vanishing in the limit where the excited-state hyperfine splitting $\rightarrow 0$ \cite{AuzTBP}.  The average splitting between the $F'$ levels that can be excited by the $^{87}$Rb $F = 1 \rightarrow F'$ and $^{85}$Rb $F = 2 \rightarrow F'$ transitions is significantly smaller than the average splitting between the $F'$ levels that can be excited by the $^{87}$Rb~$F=2 \rightarrow F'$ and $^{85}$Rb~$F=3 \rightarrow F'$ transitions, leading to the observed difference in FM NMOR rotation amplitudes.  In the following, we restrict our considerations to the $^{87}$Rb~$F=2 \rightarrow F'$ and $^{85}$Rb~$F=3 \rightarrow F'$ hyperfine components for which the FM NMOR signal amplitudes are largest, and yield the best magnetometric sensitivity.

\begin{figure*}
\centerline{\psfig{figure=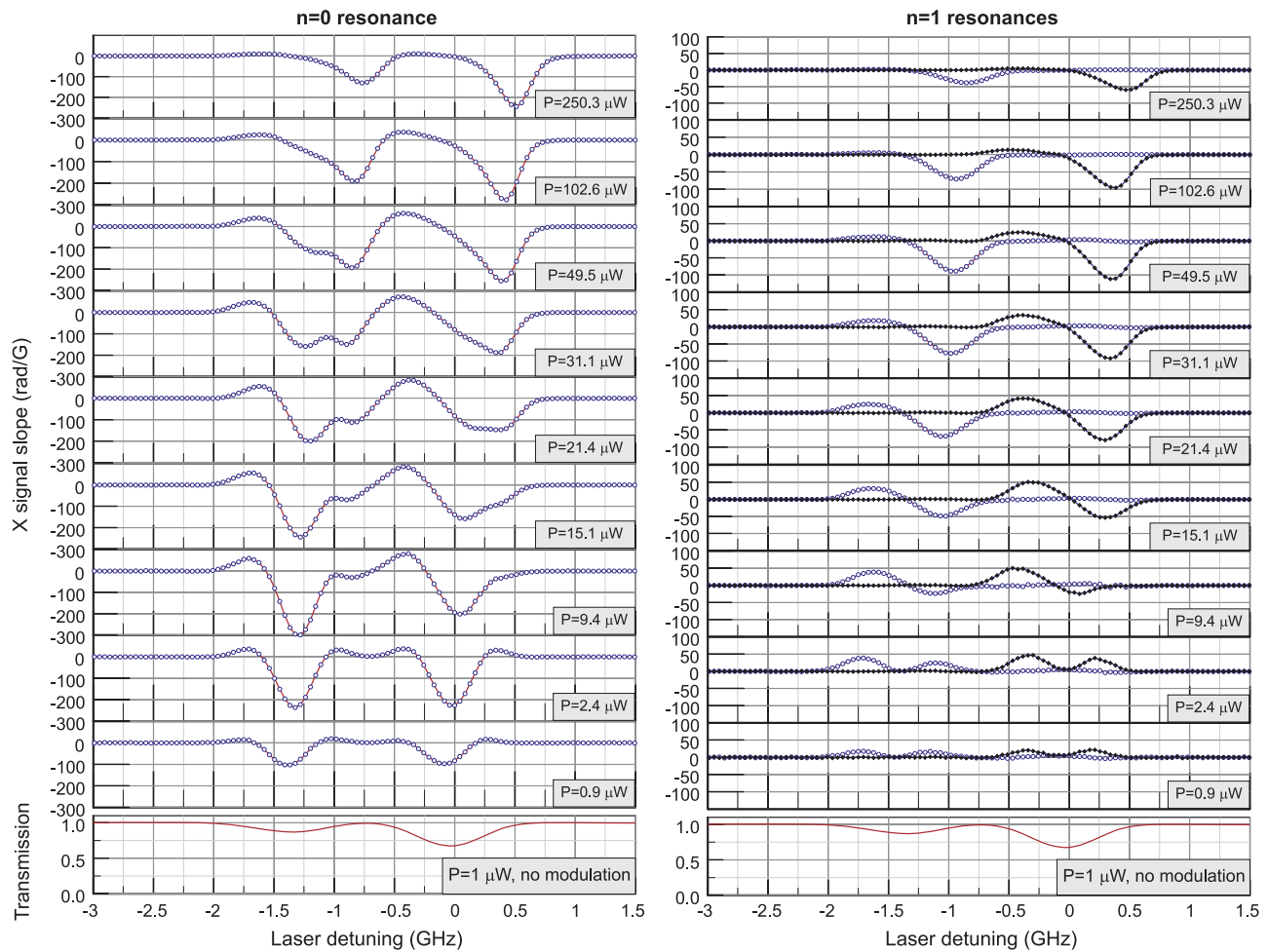,width=6.7 in}} \caption{Laser detuning dependence of the derivative of the X signal optical rotation amplitude with respect to longitudinal field ($d\varphi/dB_z$ in units of rad/G) for $n=0$ and $n=1$ resonances (left and right plots, respectively) for various light powers.  Laser modulation parameters are $\Omega_m = 500~{\rm Hz}$ and $\Delta \omega = 65~{\rm MHz}$, and the cell temperature $T = 25^\circ$C, corresponding to a vapor density of $\approx 10^{10}~{\rm atoms/cm^3}$.  For the $n=1+$ resonances shown in the plots on the right-hand side, data for $^{87}$Rb (open circles) are acquired at $B_z = 357~{\rm \mu G}$ and data for $^{85}$Rb (filled circles) are acquired at $B_z = 536~{\rm \mu G}$. Transmission spectra for light power $1~{\rm \mu W}$ (with no frequency modulation) are shown at bottom.  Only $^{87}$Rb $F=2 \rightarrow F'$ and $^{85}$Rb $F=3 \rightarrow F'$ components of the D2 transition are shown.} \label{Fig:SpectraPowerDependence}
\end{figure*}

\begin{figure}
\centerline{\psfig{figure=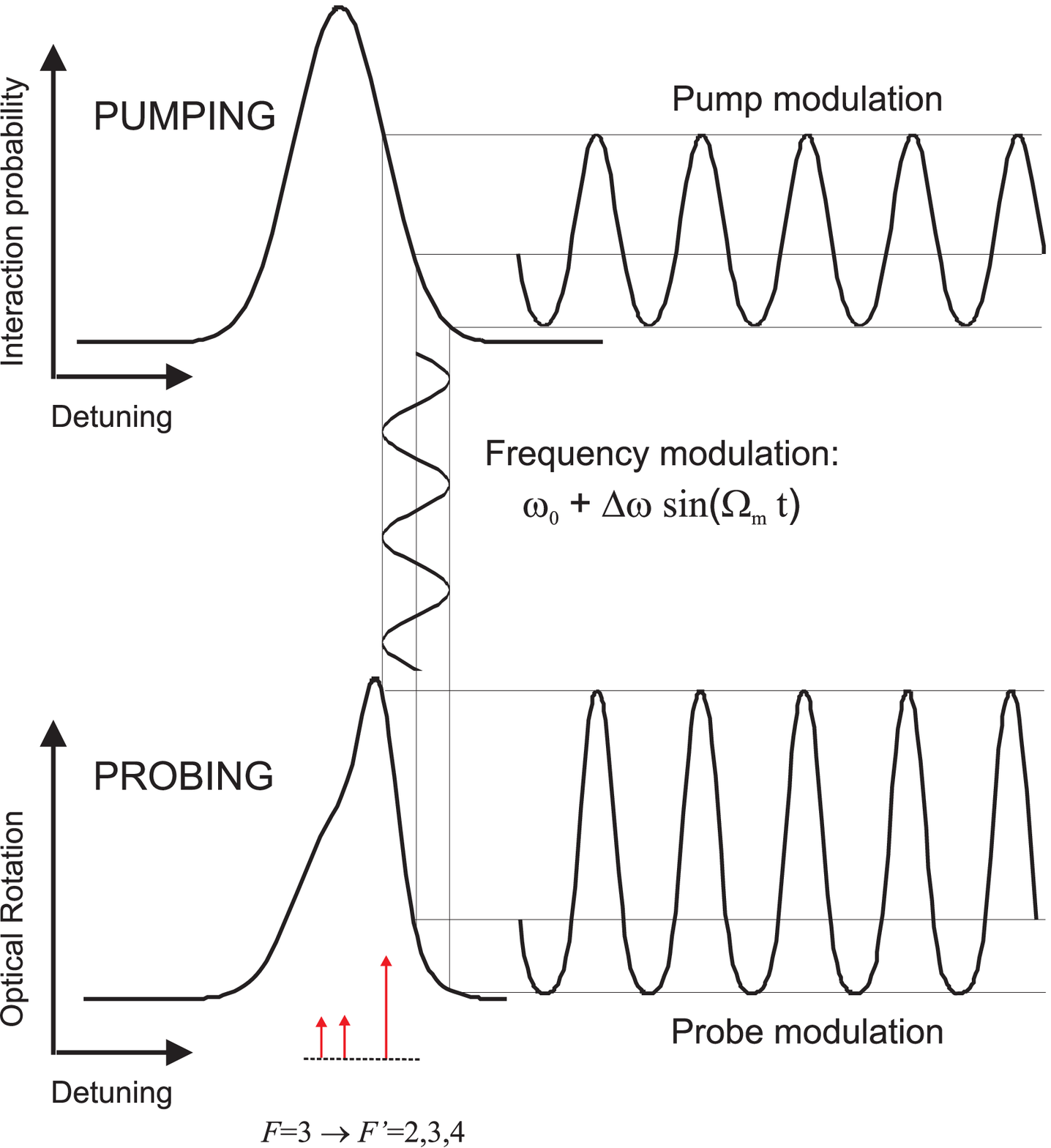,width=2.75 in}} \caption{Illustration of the time-dependent optical pumping and optical rotation generated via frequency modulation for high light power (where alignment-to-orientation is the dominant cause of optical rotation, causing enhanced rotation for $F \rightarrow F+1$ transitions with the same sign as compared to $F \rightarrow F-1, F$ transitions, see text). Arrows at the base of the probing interaction diagram indicate the approximate central frequencies and estimated relative contribution of different hyperfine components ($F \rightarrow F'$) to the overall Doppler-broadened optical rotation spectrum.} \label{Fig:High-Power-PumpProbe-Modulation}
\end{figure}

Comparing the low-light-power ($P = 2.4~{\rm \mu W}$) and high-light-power ($P = 102.6~{\rm \mu W}$) FM NMOR spectra in Fig.~\ref{Fig:LowPowerHighPowerSpectra}, a pronounced difference is observed in the shape of the spectra.  For the $n=1$ resonances, there is a reversal of the sign of the FM NMOR signals as well as change from a spectrum with two maxima for each Doppler-broadened optical resonance to a spectrum with a single large peak located at the high-frequency side of the Doppler-broadened optical resonances.  Also of note is the fact that the pronounced difference between $n=0$ and $n=1$ spectra observed in the low-light-power case disappears at higher light powers as the $n=0$ and $n=1$ spectra take on rather similar shapes.  The change in the FM NMOR spectra as a function of light power is better illustrated in Fig.~\ref{Fig:SpectraPowerDependence}, which displays a series of FM NMOR X signal slope spectra for the $^{87}$Rb~$F=2 \rightarrow F'$ and $^{85}$Rb~$F=3 \rightarrow F'$ hyperfine components at different light powers.

The change in the FM NMOR spectra as light power is increased is attributable to the phenomenon of alignment-to-orientation conversion \cite{Bud00aoc}.  The signature feature of NMOR due to AOC is a reversal of the sign of optical rotation for $F \rightarrow F+1$ transitions as light power is increased, as discussed in Ref.~\cite{Bud00aoc}, whereas the sign of optical rotation for $F \rightarrow F-1$ and $F \rightarrow F$ transitions does not change as light power is increased.  The resonant frequencies for the $F \rightarrow F+1$ components of the Doppler-broadened optical resonances occur on the high-frequency sides of the $^{87}$Rb~$F=2 \rightarrow F'$ and $^{85}$Rb~$F=3 \rightarrow F'$ transitions.  Figure~\ref{Fig:SpectraPowerDependence} shows that the sign of the FM NMOR signal on the high frequency side of the $^{87}$Rb~$F=2 \rightarrow F'$ and $^{85}$Rb~$F=3 \rightarrow F'$ hyperfine components reverses as light power is increased, suggesting that indeed the change in the spectrum at high light powers is due to AOC.  Figure~\ref{Fig:High-Power-PumpProbe-Modulation} illustrates how the change in the Doppler-broadened spectrum of NMOR without modulation (observed and discussed in Refs.~\cite{Bud00aoc,Bud00sens}), going from the dispersive shape shown in Fig.~\ref{Fig:Low-Power-PumpProbe-Modulation} to the spectrum sharply peaked on the high frequency side of the resonance shown in Fig.~\ref{Fig:High-Power-PumpProbe-Modulation}, creates the observed light-power dependence of the FM NMOR spectra.  In the high-light-power case modulation of the optical pumping interaction has strong components at the first harmonic of $\Omega_m$ at approximately the same detunings for which the optical rotation generated by the probe interaction has strong first harmonic components, causing the FM NMOR spectra for the $n=0$ and $n=1$ cases to take on similar shapes.

Of interest is the fact that even for the $n=1$ resonances, AOC creates atomic orientation parallel to $\mathbf{B}$, and thus the atomic orientation does not precess in the magnetic field.  The time-dependent optical rotation detected at the first harmonic of $\Omega_m$ is the result of modulation of the probe interaction.  Experimental schemes (for example, that described in Ref.~\cite{Hig06}) employing a pump/probe arrangement where the probe beam is unmodulated are thus generally insensitive to NMOR due to AOC.

\begin{figure}
\centerline{\psfig{figure=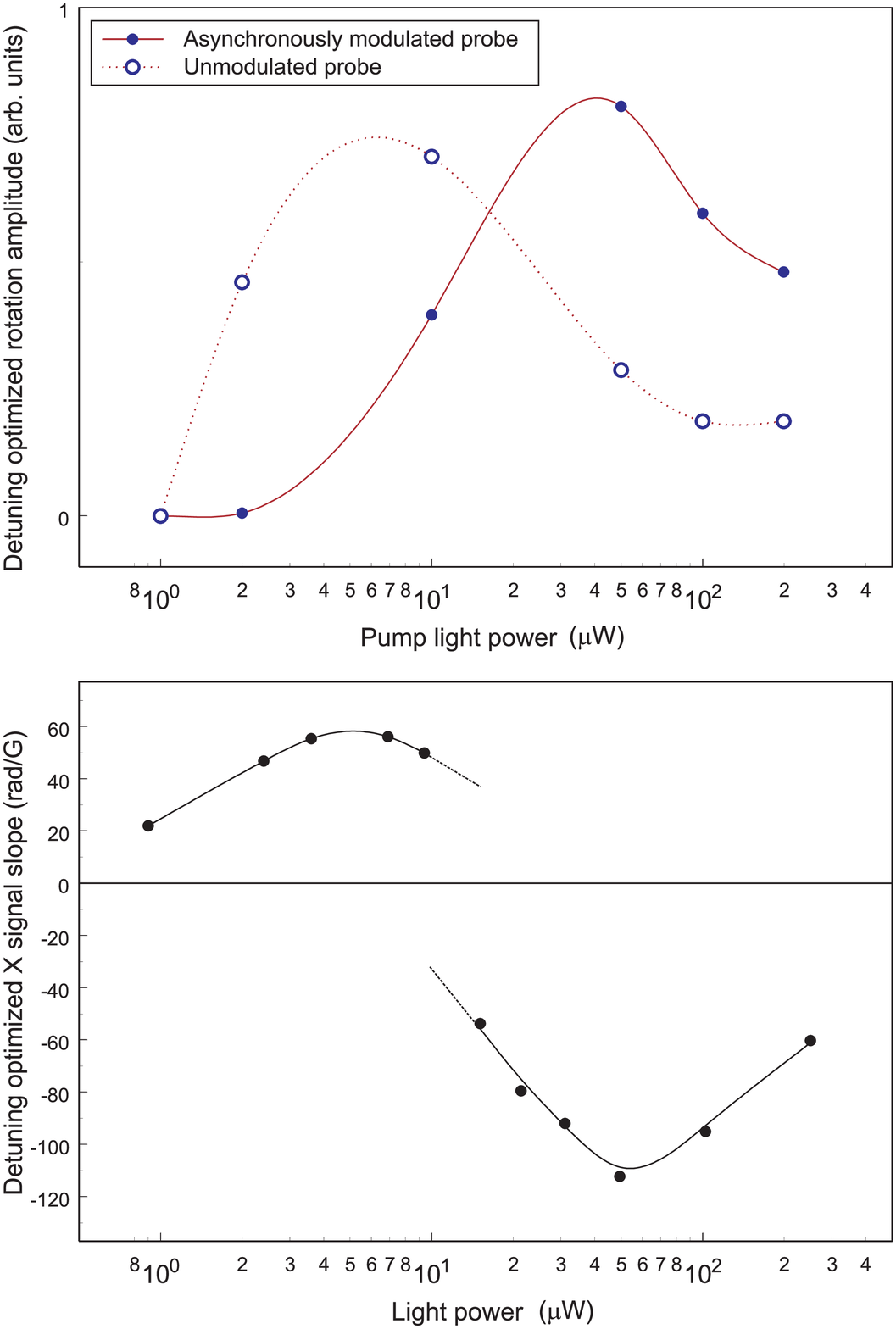,width=3.35 in}} \caption{Upper plot shows the pump-light-power dependence of the normalized, detuning-optimized amplitude of optical rotation for an unmodulated probe beam measured at the first harmonic of $\Omega_m = 2\Omega_L = 2\pi \times 500~{\rm Hz}$ (open circles, dashed line to guide the eye) and an asynchronously modulated probe beam measured at the first harmonic of $\Omega\ts{probe} = 2\pi \times 655~{\rm Hz}$ (filled circles, solid line to guide the eye).  The 795-nm probe beam measured the optical rotation spectrum for the Rb D1 line, and the probe light power in both cases was $20~\rm{\mu W}$.  The unmodulated probe beam is principally sensitive to atomic alignment transverse to $\hat{z}$ precessing at $\Omega_L$ while the asynchronously modulated probe beam is principally sensitive to static atomic orientation along $\hat{z}$.  The pump beam was tuned to the maximum of rotation for the $^{85}$Rb $F=3 \rightarrow F'$ component of the D2 line and frequency modulated at $\Omega_m = 2\pi \times 500~{\rm Hz}$ with $\Delta \omega = 2\pi \times 300~{\rm MHz}$.  The lower plot shows the dependence of the detuning-optimized amplitude of the FM NMOR X signal slope for the $^{85}$Rb $n=1$ resonance.} \label{Fig:sync-async}
\end{figure}

To verify this explanation for the light-power-dependence of the FM NMOR spectra, a second laser beam ($\lambda = 795~{\rm nm}$, resonant with the Rb D1 line) was used to independently probe the presence of static orientation and precessing alignment. The 795-nm D1 probe beam was directed collinearly with the 780-nm D2 beam along the $z$-axis of the apparatus, and a pick-off mirror was used at the output to direct the D1 probe light into a second polarimeter.  The signal from the second polarimeter was measured with a second lock-in amplifier.  The D2 beam was tuned to the maximum of the FM NMOR signal for the $^{85}$Rb $F=3 \rightarrow F'$ component of the D2 line and was frequency modulated at $\Omega_m = 2\pi \times 500~{\rm Hz}$ with $\Delta \omega = 2\pi \times 300~{\rm MHz}$.  The magnetic field was set to satisfy the $n=1$ resonance condition [Eq.~\eqref{Eq:MagFieldResonanceCondition}].  For a given power of the D2 ``pump'' beam, two different optical rotation spectra were acquired with the D1 probe beam: (1) with the probe beam unmodulated and the optical rotation spectrum demodulated by a lock-in amplifier at the first harmonic of $\Omega_m$, and (2) with the probe beam modulated at $\Omega_a = 2\pi \times 655~{\rm Hz}$ and demodulated at the first harmonic of $\Omega_a$.  Case (1) is sensitive only to atomic polarization that is precessing in $\mathbf{B}$ at a rate $2\Omega_L = \Omega_m$, and thus is associated with atomic alignment transverse to $\mathbf{B}$.  This is because the polarization of the unmodulated probe must acquire its time-dependence at frequency $\Omega_m$ from the dynamics of the atomic spins. Case (2), where the probe beam undergoes asynchronous modulation [since it is far from the resonance condition, Eq.~\eqref{Eq:MagFieldResonanceCondition}], is sensitive to static polarization moments that generate optical rotation, in particular orientation along $z$.  The time-dependence of the Rb atomic spin polarization has no detectable frequency component at $\Omega_a$, so the observed time-dependent optical rotation at frequency $\Omega_a$ must result from the periodicity of the interaction of the probe light with static atomic orientation along $z$ (which causes optical rotation via circular birefringence when the probe light is resonant with a component of the Doppler-broadened optical transition).

\begin{figure*}
\centerline{\psfig{figure=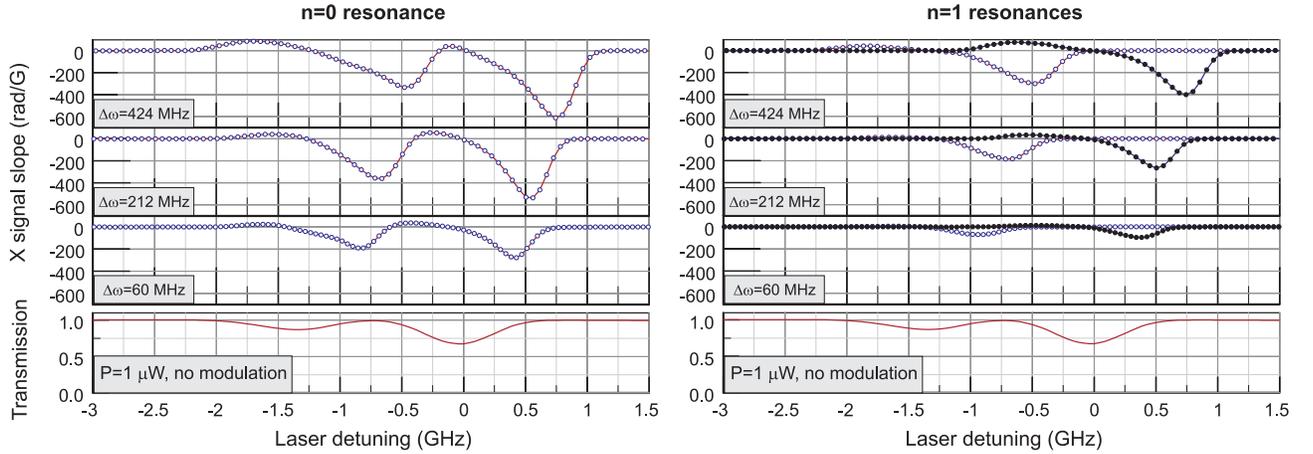,width=6.7 in}} \caption{Laser detuning dependence of the derivative of the X signal optical rotation amplitude with respect to longitudinal field ($d\varphi/dB_z$ in units of rad/G) for $n=0$ and $n=1$ resonances (left and right plots, respectively) for various modulation amplitudes $\Delta \omega$.  Laser modulation parameters are $\Omega_m = 500~{\rm Hz}$ and light power $= 100~{\rm \mu W}$, and the cell temperature $T = 20^\circ$C, corresponding to a vapor density of $\approx 4 \times 10^{9}~{\rm atoms/cm^3}$.  For the $n=1+$ resonances shown in the plots on the right-hand side, data for $^{87}$Rb (open circles) are acquired at $B_z = 357~{\rm \mu G}$ and data for $^{85}$Rb (filled circles) are acquired at $B_z = 536~{\rm \mu G}$. Transmission spectra for light power $1~{\rm \mu W}$ (with no frequency modulation) are shown at bottom.  Only $^{87}$Rb $F=2 \rightarrow F'$ and $^{85}$Rb $F=3 \rightarrow F'$ components of the D2 transition are shown.} \label{Fig:SpectraModFreqAmpDependence}
\end{figure*}

Figure~\ref{Fig:sync-async} shows the results of this measurement.  The lower plot of Fig.~\ref{Fig:sync-async} displays the detuning-optimized FM NMOR X signal slope for the $^{85}$Rb $n=1$ resonance obtained from the data presented in Fig.~\ref{Fig:SpectraPowerDependence}. At low light powers ($P \approx 0 - 15~{\rm \mu W}$), the maximum FM NMOR X signal slope is positive and occurs at the low-frequency side of the Doppler-broadened $F=3 \rightarrow F'$ optical resonance; at high light powers ($P \gtrsim 15~{\rm \mu W}$), the maximum FM NMOR X signal slope is negative and occurs at the high-frequency side of the Doppler-broadened $F=3 \rightarrow F'$ optical resonance.  As seen in the lower plot of Fig.~\ref{Fig:sync-async}, in the low-light-power regime, believed to be associated with alignment precession, the FM NMOR amplitude peaks at $P \approx 6~{\rm \mu W}$, while in the high-light-power regime, believed to be associated with AOC, the FM NMOR amplitude peaks at $P \approx 50~{\rm \mu W}$.  The upper plot of Fig.~\ref{Fig:sync-async} displays the detuning-optimized optical rotation amplitude for both an unmodulated D1 probe measured at the first harmonic of $\Omega_m$ (sensitive to precessing alignment) and an asynchronously modulated D1 probe measured at the first harmonic of $\Omega_a$ (sensitive to static orientation along $\mathbf{k}$, which is parallel to $\mathbf{B}$) as a function of the D2 pump light power.  The optical rotation amplitude for the unmodulated probe peaks at $P \approx 6~{\rm \mu W}$ while the optical rotation for the asynchronously modulated probe peaks at $P \approx 50~{\rm \mu W}$, confirming the explanation for the light-power-dependence of the FM NMOR spectra in terms of AOC.

\section{Modulation amplitude dependence of FM NMOR spectra}
\label{Sec:ModAmp}

Another important experimental parameter to be considered in magnetometric sensitivity optimization is the frequency modulation amplitude $\Delta \omega$.  Figure~\ref{Fig:SpectraModFreqAmpDependence} shows the FM NMOR X signal slope spectra for the $^{87}$Rb $F=2 \rightarrow F'$ and $^{85}$Rb $F=3 \rightarrow F'$ components of the D2 transition for representative values of $\Delta \omega$.   When $\Delta \omega \ll \Gamma_D$, where $\Gamma_D \approx 2 \pi \times 330~{\rm MHz}$ is the Doppler width, sinusoidal frequency modulation with the light detuned to the wing of a Doppler-broadened optical resonance produces nearly sinusoidal modulation of the light-atom interaction probability.  When $\Delta \omega \gtrsim \Gamma_D$, sinusoidal frequency modulation generally produces non-sinusoidal modulation of the light-atom interaction probability, and thus modulation of the pump and probe interactions at $\Omega_m$ and higher harmonics of $\Omega_m$.  This effect can produce some distortion of the FM NMOR spectra at large $\Delta \omega$.  Additionally, as $\Delta \omega$ increases, the FM NMOR spectra are broadened since they are the convolution of the frequency modulated laser spectrum and the Doppler broadened optical resonance.   The detuning-optimized FM NMOR amplitude reaches a maximum at $\Delta \omega \sim \Gamma_D$ for each hyperfine component, which is expected since $\Delta \omega \sim \Gamma_D$ yields the maximum possible modulation of the light-atom interaction probability.

\section{Magnetometric sensitivity}

The sensitivity of any atomic magnetometer is fundamentally limited by atomic shot noise: the unavoidable quantum uncertainty in the measurement of the atomic spin projection along a spatial axis.  The fundamental atomic-shot-noise-limited sensitivity is given by Eq.~\eqref{Eq:FundamentalSensitivity-atoms} (techniques such as spin-squeezing and quantum non-demolition measurements only offer gains in magnetometric sensitivity for time scales short compared to $\gamma\ts{rel}^{-1}$, ultimately a measurement integrated over a time scale long compared to $\gamma\ts{rel}^{-1}$ obtains the sensitivity described by Eq.~\eqref{Eq:FundamentalSensitivity-atoms}, see Ref.~\cite{Auz06}).  For an all-optical atomic magnetometer, as considered here, if optimal efficiency of optical pumping and probing of atomic spins is achieved, magnetometric sensitivity should be limited in equal parts by the atomic shot noise and the photon shot noise \cite{Auz06}.  The fundamental atom and photon shot-noise-limited magnetometric sensitivity [$\sqrt{2}$ times the value from Eq.~\eqref{Eq:FundamentalSensitivity-atoms}] under our typical experimental conditions (at a cell temperature of $\approx 20^\circ$C yielding an Rb vapor density of $\approx 4 \times 10^{9}~{\rm atoms/cm^3}$, zero-light-power-extrapolated relaxation rate $\gamma\ts{rel}(0) \approx 1.3~{\rm Hz}$, and vapor cell radius $R \approx 2.5~{\rm cm}$) is:
\begin{align}
^{87}{\rm Rb}:&~~~\sqrt{2} \times \delta B\ts{SNL} \approx 5 \times 10^{-12}~{\rm G}/\sqrt{\rm Hz}~ \nonumber \\
^{85}{\rm Rb}:&~~~\sqrt{2} \times \delta B\ts{SNL} \approx 5 \times 10^{-12}~{\rm G}/\sqrt{\rm Hz}~, \nonumber
\end{align}
where we account for both the natural isotopic abundance and the different gyromagnetic ratios of the two isotopes.  This fundamental sensitivity limit is entirely derived from the properties of the paraffin-coated cell and the Rb atoms, and has nothing to do with the specific experimental technique of FM NMOR (except for the fact that in FM NMOR, in contrast to a SERF magnetometer \cite{All02,Kom03}, spin-exchange collisions contribute to relaxation).  For consistency with previous measurements of magnetometric sensitivity \cite{Bud00sens}, we define the relationship between an integration time of $\tau = 1~{\rm s}$ and a bandwidth of $1~{\rm Hz}$ to be $\sqrt{\tau = 1~{\rm s}} \leftrightarrow \sqrt{\pi}~{\rm Hz}^{-1/2}$.  (This means that in order to achieve a 1-Hz bandwidth, the measurement needs to proceed for a time $1/\pi$~s. Note that another common convention is where a 1-Hz bandwidth corresponds to a 0.5-s measurement.  There is thus relatively little numerical difference between the conventions.)

\begin{figure*}
\centerline{\psfig{figure=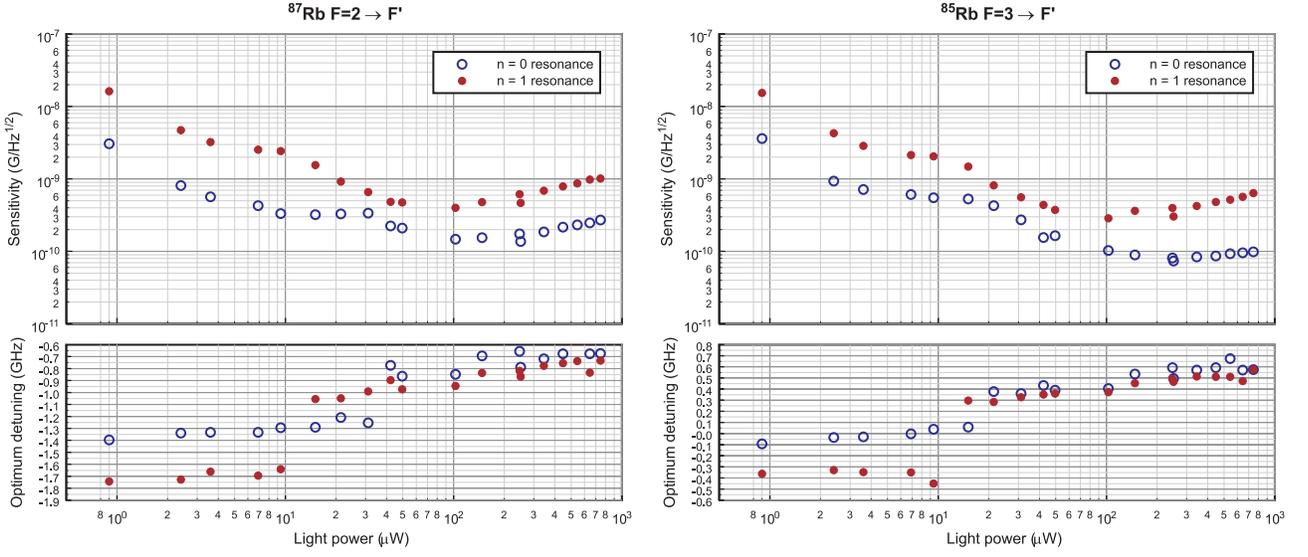,width=6.7 in}} \caption{Upper plots show laser-detuning-optimized shot-noise-projected sensitivity as a function of incident light power for both the $n=0$ and $n=1$ X signals.  Lower plots show the relative detuning at which the optimum sensitivity is achieved (zero detuning is defined to be at the center of the Doppler-broadened $^{85}$Rb $F=3 \rightarrow F'$ transition --- see, for example, Fig.~\ref{Fig:LowPowerHighPowerSpectra}).  Data acquired for cell temperature $\approx 21^\circ$C, corresponding to a vapor density of $\approx 4 \times 10^9~{\rm atoms/cm^3}$, $\Omega_m = 2\pi \times 500~{\rm Hz}$, and $\Delta \omega = 65~{\rm MHz}$.} \label{Fig:Sensitivity-vs-Power}
\end{figure*}

\begin{figure*}
\centerline{\psfig{figure=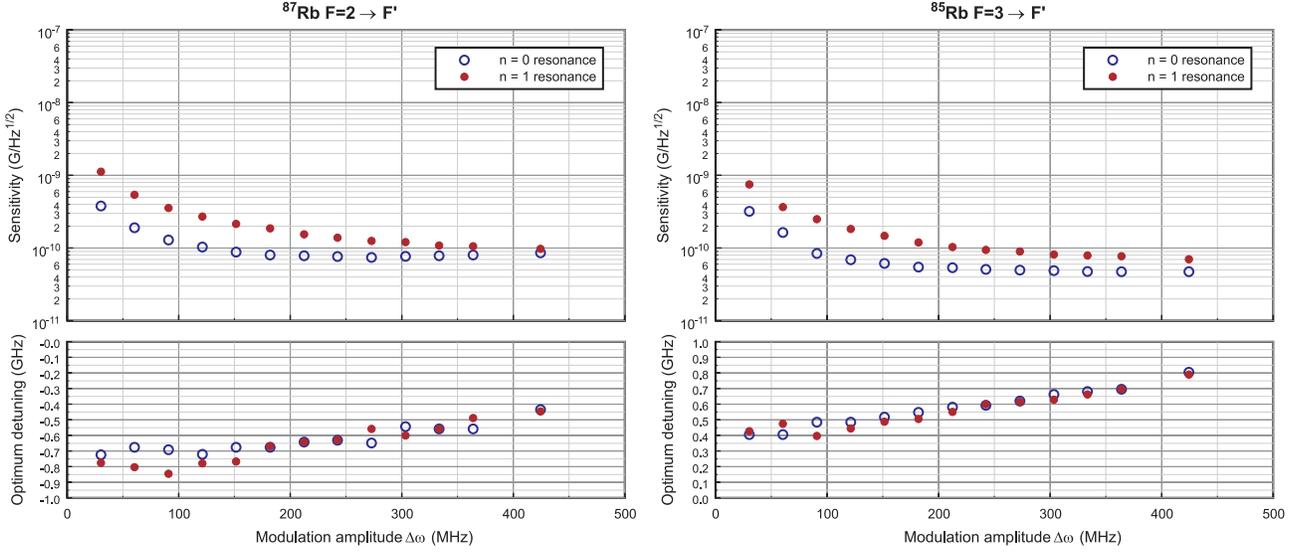,width=6.7 in}} \caption{Upper plots show laser-detuning-optimized shot-noise-projected sensitivity as a function of modulation amplitude $\Delta \omega$ for both the $n=0$ and $n=1$ X signals.  Lower plots show the relative detuning at which the optimum sensitivity is achieved (zero detuning is defined to be at the center of the Doppler-broadened $^{85}$Rb $F=3 \rightarrow F'$ transition --- see, for example, Fig.~\ref{Fig:LowPowerHighPowerSpectra}).  Data acquired for cell temperature $\approx 21^\circ$C, corresponding to a vapor density of $\approx 4 \times 10^9~{\rm atoms/cm^3}$, $\Omega_m = 2\pi \times 500~{\rm Hz}$, and incident laser light power $= 100~{\rm \mu W}$.} \label{Fig:Sensitivity-vs-ModAmp}
\end{figure*}

The overarching goal of the present study is to determine the experimental parameters for which FM NMOR achieves optimal magnetometric sensitivity and how close an FM-NMOR-based magnetometer comes to the fundamental sensitivity limit for an all-optical atomic magnetometer.  In principle, the sensitivity of a magnetometer can be determined via direct measurement in a magnetic field environment with magnetic field stability exceeding the magnetometric sensitivity, or via a differential measurement using identical magnetometers measuring the field in approximately the same region of space.  For parameter optimization, however, it is more efficient to determine the response of the the optical signal to a known change in the magnetic field and estimate the sensitivity based on the light power reaching the detector.  This approach separates the parameter optimization problem from the potentially difficult issues connected with eliminating technical sources of noise limiting the practical magnetometric sensitivity.  Thus in Figs.~\ref{Fig:Sensitivity-vs-Power} and \ref{Fig:Sensitivity-vs-ModAmp} we present determinations of the shot-noise-projected (SNP) magnetometric sensitivity,
\begin{align}
\delta B\ts{SNP} = \prn{\frac{d\varphi}{dB}}^{-1} \delta \varphi~,
\label{Eq:shot-noise-projected-sensitivity}
\end{align}
where $d\varphi/dB$ is the slope of the FM NMOR X signal (with respect to the applied longitudinal magnetic field) and $\delta \varphi$ is the photon-shot-noise-projected sensitivity of the polarimeter given the detected light power.  Independent measurements of the polarimeter sensitivity for typical experimental conditions have confirmed operation within a factor of $\sim 2$ of the shot-noise limit at modulation frequencies $\gtrsim 1~{\rm kHz}$.  Note that this parametrization ignores the Y signal, which for the $n=1$ case is potentially equally sensitive to the magnetic field, and thus we anticipate that the true magnetometric sensitivity for an optimal measurement scheme involving both X and Y $n=1$ FM NMOR signals should be close to twice the value presented here (the factor of 2 rather than $\sqrt{2}$ is because the two measurements should be strongly correlated as they are derived from a single time-dependent rotation signal).  This means that our measurements indicate no difference between the magnetometric sensitivity of the $n=0$ and $n=1$ resonances.  A previous study \cite{Bud00sens} showed that NMOR in a paraffin-coated cell using unmodulated light can achieve a shot-noise-projected sensitivity close to the fundamental limit, and it has been proposed \cite{Bud02} that FM NMOR should in principle be able to achieve a similar magnetometric sensitivity.

Figure \ref{Fig:Sensitivity-vs-Power} presents the detuning-optimized SNP magnetometric sensitivity determined from Eq.~\eqref{Eq:shot-noise-projected-sensitivity} as a function of light power for the $^{87}$Rb $F=2 \rightarrow F'$ and $^{85}$Rb $F=3 \rightarrow F'$ components of the D2 transition for a modulation amplitude of $\Delta \omega = 65~{\rm MHz}$.  The lower plots show the detuning (using the scale employed in Figs.~\ref{Fig:LowPowerHighPowerSpectra}, \ref{Fig:SpectraPowerDependence}, and \ref{Fig:SpectraModFreqAmpDependence}) at which optimum sensitivity is achieved.  Note the jump in the optimal detuning occurring at the transition between the low-light-power regime where optical rotation results from precession of atomic alignment and the high-light-power regime where optical rotation results primarily from AOC.  Figure~\ref{Fig:Sensitivity-vs-ModAmp} presents the detuning-optimized SNP magnetometric sensitivity as a function of the modulation amplitude $\Delta \omega$ at a light power of $100~{\rm \mu W}$.  The best SNP sensitivity for both isotopes corresponds to
\begin{align}
^{87}{\rm Rb}:&~~~\delta B\ts{SNP} \approx 7 \times 10^{-11}~{\rm G}/\sqrt{\rm Hz}~ \nonumber \\
^{85}{\rm Rb}:&~~~\delta B\ts{SNP} \approx 5 \times 10^{-11}~{\rm G}/\sqrt{\rm Hz}~, \nonumber
\end{align}
roughly one order of magnitude away from the fundamental limit under our experimental conditions.  Taking into account the combined sensitivity of the X and Y $n=1$ signals, the $n=0$ and $n=1$ signals achieve approximately equal magnetometric sensitivity.

Several factors may contribute to the reduced SNP magnetometric sensitivity of FM NMOR as compared to the fundamental sensitivity limit.  One factor is the optical pumping efficiency: in order to reach the fundamental sensitivity limit, all $N$ atoms in the vapor cell must participate in the measurement.  A significant fraction of atoms at the light powers where optimal SNP sensitivity is achieved are optically pumped out of the ground-state hyperfine level with which the laser light is resonant (roughly 80\% of the equilibrium population, according to estimates based on optical transition saturation parameters \cite{OurBook} and observed transmission spectra) and into the other ground-state hyperfine level where they do not participate in the measurement, thereby reducing the potential sensitivity (by a factor $\sim 2$).  Optimal contrast in the light-atom interaction probability is obtained for $\Delta \omega \sim \Gamma_D$, so even though light tuned to the high frequency wing of the Doppler-broadened $^{87}$Rb $F=2 \rightarrow F'$ and $^{85}$Rb $F=3 \rightarrow F'$ optical resonances interacts primarily with $F \rightarrow F+1$ cycling transitions, losses to the unobserved ground-state hyperfine level are unavoidable due to the effectively broad spectral profile of the laser light due to frequency modulation.  This loss mechanism may be overcome with the use of a re-pump laser resonant with the unobserved ground-state hyperfine level, an approach we plan to implement in the near future.  Additionally, the effectively broad spectral profile of the laser necessarily means that multiple hyperfine components of the Doppler-broadened optical transitions are excited, which can reduce the FM NMOR amplitude due to cancelation between the contributions of different hyperfine components to the optical pumping and probing of alignment \cite{AuzTBP}.  Another factor degrading the SNP sensitivity is light-induced relaxation of the atomic polarization: under the experimental conditions where optimum SNP sensitivity is achieved, $\gamma\ts{rel} \sim 2\pi \times 10~{\rm Hz}$ due to light-induced relaxation.  As can be seen from Eq.~\eqref{Eq:FundamentalSensitivity-atoms}, this reduces the potential sensitivity of the magnetometer by a factor of $\sim 3$.  It appears that the combination of these factors accounts for most of the gap between the measured SNP magnetometric sensitivity of FM NMOR and the fundamental limit.

In a previous study \cite{Bud00sens} of NMOR using unmodulated light, where the SNP sensitivity was found to be on the order of the fundamental limit, the optimum SNP magnetometric sensitivity for the D2 line was achieved when the laser light was detuned far to the high frequency wing of the Doppler-broadened $^{85}$Rb $F=3 \rightarrow F'$ optical resonance.  Since the unmodulated light was narrow band ($\sim 1~{\rm MHz}$), the light selectively interacted with the $F=3 \rightarrow F'=4$ cycling transition, significantly reducing optical pumping to the unobserved $F=2$ ground-state hyperfine level, and because the light was significantly detuned from the center of the Doppler-broadened optical resonance, the light-induced relaxation was reduced to on the order of the zero-light-power-extrapolated relaxation rate \cite{Bud98}.  Thus both identified loss factors for FM NMOR are significantly improved upon using NMOR with unmodulated light.

Improvement in the fundamental sensitivity by increasing the number of alkali atoms $N$ [see Eq.~\eqref{Eq:FundamentalSensitivity-atoms}] is limited in practice \cite{endnote2} at high densities because $\gamma\ts{rel}$ tends to scale with $N$ due to spin-exchange collisions \cite{Bud00sens} (there is also evidence for temperature-dependent wall relaxation \cite{Gra05} that may similarly limit possible improvement in sensitivity by increasing $N$ via cell heating).  However, since light broadening rather than collisional relaxation dominates $\gamma\ts{rel}$ for the experimental conditions where optimum SNP sensitivity is obtained in our case, some improvement in sensitivity can be obtained by increasing $N$ via heating the cell.

\begin{figure}
\centerline{\psfig{figure=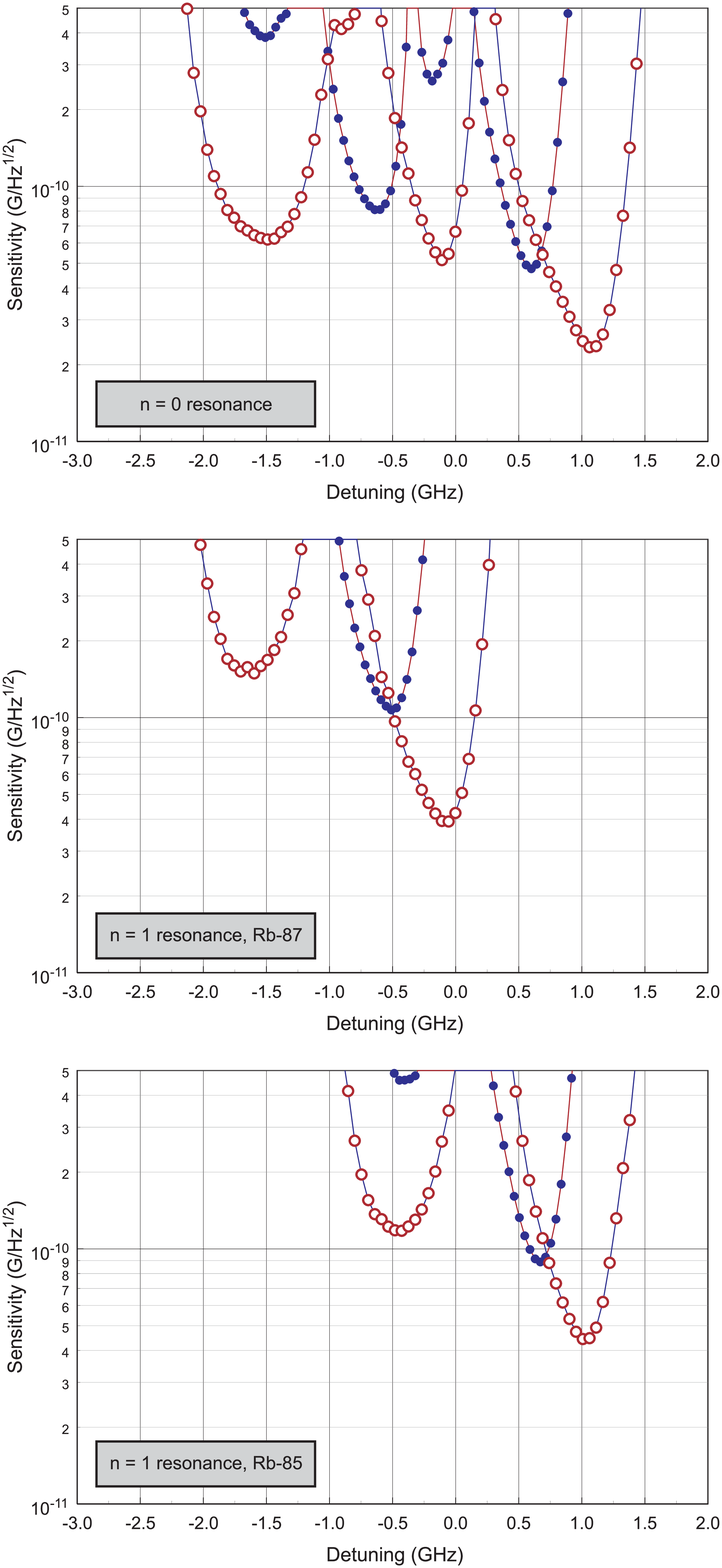,width=3.35 in}} \caption{Magnetometric sensitivity as a function of detuning from the center of the Doppler-broadened $^{85}$Rb $F=3 \rightarrow F'$ resonance.  Filled circles (blue) show sensitivity for cell temperature $T \approx 21^\circ$C and Rb vapor density $\approx 4 \times 10^9~{\rm atoms/cm^3}$, incident laser light power $= 100~{\rm \mu W}$, modulation amplitude $\Delta \omega = 2\pi \times 300~{\rm MHz}$, and $\Omega_m = 2\pi \times 500~{\rm Hz}$. Open circles (red) show sensitivity for cell temperature $T \approx 35^\circ$C and Rb vapor density $\approx 29 \times 10^9~{\rm atoms/cm^3}$, incident laser light power $= 317~{\rm \mu W}$, modulation amplitude $\Delta \omega = 2\pi \times 545~{\rm MHz}$, and $\Omega_m = 2\pi \times 500~{\rm Hz}$.  Note that the SNP magnetometric sensitivity spectra reflect differences between the FM NMOR spectra at different Rb vapor densities, light powers, and modulation amplitudes.} \label{Fig:HighDensity}
\end{figure}

To demonstrate this possibility, we increased $N$ by heating the cell with warm air blown into the innermost shield.  Systematic measurements proved difficult because the cell temperature obtained by this method was neither particularly stable nor easy to adjust, and we are presently constructing a new cell temperature control and stabilization system for future experiments.  Nonetheless, it is of interest to note the improvement in the SNP magnetometric sensitivity obtained at higher $N$ shown in Fig.~\ref{Fig:HighDensity}; with the cell temperature $T \approx 35^\circ$C and Rb vapor density $\approx 29 \times 10^9~{\rm atoms/cm^3}$, SNP magnetometric sensitivity of $\delta B\ts{SNP} \approx 2 \times 10^{-11}~{\rm G}/\sqrt{\rm Hz}$ is obtained for nearly optimal laser light power and frequency-modulation amplitude.

\section{Conclusion}

We have carried out a systematic optimization of the magnetometric sensitivity of nonlinear magneto-optical rotation with frequency-modulated light (FM NMOR) for the Rb D2 line with respect to light power, light detuning, and frequency-modulation amplitude.  We have found that the best shot-noise-projected (SNP) magnetometric sensitivity under optimum conditions for a 5-cm-diameter paraffin-coated cell at temperature $T \approx 20^\circ$C (corresponding to a Rb vapor density of $4 \times 10^9~{\rm atoms/cm^s}$) is $\delta B\ts{SNP} \approx 5 \times 10^{-11}~{\rm G}/\sqrt{\rm Hz}$, around an order-of-magnitude above the fundamental quantum limit for an ideal atomic magnetometer.  Two factors reducing the magnetometric sensitivity were identified: (1) optical pumping to the unobserved ground-state hyperfine level, (2) light-induced relaxation of the optically pumped ground-state atomic polarization.  In future studies, we plan to investigate the use of a re-pump laser to address the reduction of magnetometric sensitivity due to factor (1).  We have also studied how increasing the Rb vapor density through cell heating can improve magnetometric sensitivity, and have observed a SNP magnetometric sensitivity of $\delta B\ts{SNP} \approx 2 \times 10^{-11}~{\rm G}/\sqrt{\rm Hz}$ for a cell temperature of $T \approx 35^\circ$C (corresponding to a Rb vapor density of $29 \times 10^9~{\rm atoms/cm^s}$).

Detailed study of the FM NMOR spectra as a function of light power and frequency-modulation amplitude were carried out.  The low-light-power ($P \lesssim 15~{\rm \mu W}$) spectra for the $n=0$ FM NMOR resonance (centered around $\Omega_L = 0$, where $\Omega_L$ is the Larmor frequency) were found to differ significantly from the spectra of the low-light-power $n=1$ FM NMOR resonance (centered around $\Omega_L = \Omega_m/2$, where $\Omega_m$ is the light modulation frequency).  This difference was attributed to the requirement of synchronous optical pumping for the $n=1$ resonances and the absence of such a requirement for the $n=0$ resonance.  A pronounced change in the FM NMOR spectra was observed as a function of light power: this change was shown through auxiliary experiments to be due to the phenomenon of alignment-to-orientation conversion (AOC, see Ref.~\cite{Bud00aoc}), where ac Stark shifts due to the optical electric field combine with the Zeeman shifts due to the applied magnetic field $\mathbf{B}$ to generate orientation along $\mathbf{B}$ from the optically pumped ground-state atomic alignment whose axis is initially parallel to the light polarization.  Alignment-to-orientation conversion significantly enhances magnetometric sensitivity for FM NMOR at high light powers ($\gtrsim 15~{\rm \mu W}$).

The immediate future goals of our research are to complete a similar optimization of the magnetometric sensitivity of FM NMOR for the Rb D1 line and to study the Allan variance of an FM-NMOR-based magnetometer with the goal of achieving the SNP magnetometric sensitivity in practice.  We also plan to develop a gyroscope sensor based on simultaneous measurement of FM NMOR $n=1$ resonances for both Rb isotopes.

The long-term goal of our research program is to use the techniques of FM-NMOR-based magnetometry for tests of fundamental physics.  The techniques used in atomic magnetometry for precise measurement of atomic spin precession and Zeeman shifts can also be used to search for anomalous Zeeman shifts not associated with magnetic fields.  Our best SNP magnetometric sensitivity translates into a sensitivity to atomic spin precession frequencies of $\approx 10~{\rm \mu Hz/\sqrt{Hz}}$ or to Zeeman sublevel shifts of $\approx 4 \times 10^{-20}~{\rm eV/\sqrt{Hz}}$).

We are presently initiating a new search for a long-range coupling between Rb nuclear spins and the mass of the Earth.  In the envisioned experiment, the electron spins are employed as co-magnetometers for the nuclear spins, allowing precise control over any magnetic-field-related systematic effects.  If interpreted as a limit on a spin-gravity interaction of the form $\mathbf{I}\cdot\mathbf{g}$ between nuclear spins $\mathbf{I}$ and the gravitational field of the Earth $\mathbf{g}$, a simultaneous measurement of the spin precession of $^{85}$Rb and $^{87}$Rb at the SNP sensitivity determined in the present work (assuming an integration time of $t \sim 10^6~{\rm s}$, which leads to a statistics-limited sensitivity of $\sim 10^{-8}~{\rm Hz}$ to spin precession or $\sim 4 \times 10^{-23}~{\rm eV}$ to Zeeman shifts) would improve on the present best experimental limit \cite{Ven92} on the coupling of the proton spin to gravity by over two orders of magnitude and match the present best experimental limit \cite{Hec08} on spin-gravity couplings in general.

Application of FM NMOR techniques to other problems in fundamental physics, such as searches for parity-violating and time-reversal-invariance-violating permanent electric dipole moments \cite{Reg02,Gri09}, continue to be considered \cite{Kim09}, and FM-NMOR-based magnetometers have already found numerous applications ranging from nuclear magnetic resonance experiments \cite{Yas04,Cra08,Led09} and magnetic resonance imaging studies \cite{Xu06a,Xu06b,Xu08} to geophysical field measurements \cite{Aco06} and magnetic particle detection \cite{Xu06c}.  We anticipate that the sensitivity optimization carried out in the present work will expand and improve the capabilities of FM-NMOR-based all-optical atomic magnetometers to address a wide range of problems in fundamental and applied physics.

\acknowledgments

We are deeply indebted to Dmitry Budker and Micah Ledbetter for numerous useful scientific discussions, to Valeriy Yashchuk for the design of the magnetic shield system, to Misha Balabas for manufacturing the paraffin-coated cells, to Valentin Dutertre for the design and modeling of the magnetic field coils, to Mohammad Ali for his outstanding technical work in support of this experiment, and to HinYan Chan for work on the laser frequency stabilization apparatus.  We thank Khoa Nguyen for his contributions to the early stages of this work, and Ian Lacey and Sahar Muhsin for their important work on the two-laser-beam measurement of synchronous and asynchronous magneto-optical rotation that confirmed the explanation of the light-power-dependent FM NMOR spectra in terms of alignment-to-orientation conversion.

We sincerely appreciate the work of Contra Costa Community College students Morgan Jacobs, Kristopher Pohlman, and Dylan Gorman and Foothill Community College student Jennifer Strange who assembled the magnetic shield system.

This work was supported in part by the National Science Foundation under grant PHY-0652824 and Faculty Support Grants from California State University - East Bay (CSUEB).  Any opinions, findings and conclusions or recommendations expressed in this material are those of the authors and do not necessarily reflect those of the National Science Foundation or CSUEB.


\begin{thebibliography}{99}
\bibitem{endnote1}
There are contributions to the X signal from $n=0$ FM NMOR resonances with widths determined by the transit rate of atoms through the laser beam \cite{Bud98} and the natural width of of the optical transition (Bennett structure effects \cite{Bud02holeburning}) which generate a non-zero background signal in the X channel at the frequency of the $n=1$ resonance, see Ref.~\cite{Bud02review}.
\bibitem{endnote2}
This limit, as previously discussed, is notably overcome in a SERF magnetometer \cite{All02,Kom03,Sha07,Led08} which operates under conditions where rapid spin-exchange causes atomic polarization in the two ground-state hyperfine levels to become strongly correlated and spin-exchange relaxation is effectively eliminated.
\end{thebibliography}
\end{document}